\documentclass[prd,twocolumn,showpacs,preprintnumbers,floatfix,superscriptaddress,10pt]{revtex4}
\usepackage[mathscr]{eucal}
\usepackage[dvips]{color}
\usepackage[dvips]{graphicx}
\usepackage{epsf}
\usepackage{bm}
\usepackage{amssymb}
\usepackage{amsmath}
\usepackage[normalem]{ulem}

\usepackage[normalem]{ulem}  % \sout{old text} for strikeout

\newcommand{\be}{\begin{equation}}

\newcommand{\ee}{\end{equation}}
\newcommand{\bea}{\begin{eqnarray}}
\newcommand{\eea}{\end{eqnarray}}

\newcommand{\de}{\partial}
\newcommand{\fourier}[1]{\int \frac{d^4#1}{(2\pi)^4}}
\newcommand{\intspace}[1]{\int d^4 {#1}\,}
\newcommand{\ha}{\frac{1}{2}}
\newcommand{\rr}{{\bf r}}
\newcommand{\q}[2]{ {\bf q}_{#1}^{#2}}
\newcommand{\qia}{ {\bf q}_{I}^{a}}
\newcommand{\setq}[2]{\{{\bf q}_{#1}^{#2}\}}

\newcommand{\unitq}[2]{\hat{{\bf q}}_{#1}^{#2}}
\newcommand{\unitqia}{\hat{{\bf q}}_{I}^{a}}
\newcommand{\hatq}[2]{\{\hat{{\bf q}}_{#1}^{#2}\}}

\newcommand{\fiak}[3]{\phi_{#1}^{#2}(#3)}

\newcommand{\dm}[1]{\delta\mu_{#1}}
\newcommand{\dmi}{\delta\mu_{I}}
\newcommand{\kpe}{{\bf{k}}_{\bot a}}
\newcommand{\kpa}{{\bf{k}}_{|| a}}
\newcommand{\coleps}{\varepsilon_{I\alpha\beta}}
\newcommand{\flaeps}{\varepsilon_{Iij}}
\newcommand{\vk}{{\bf{k}}}

\newcommand{\vu}{{\bf u}}
\newcommand{\vv}{\hat{{\bf v}}}
\newcommand{\cf}{\alpha i,\beta j}
\newcommand{\bp}{\bar{\psi}}
\newcommand{\pll}{\psi_{\langle}}
\newcommand{\bpl}{\bar{\psi}_{\langle}}
\newcommand{\pg}{\psi_{\rangle}}
\newcommand{\bpg}{\bar{\psi}_{\rangle}}
\newcommand{\cll}{\chi_{\langle}}
\newcommand{\bcl}{\bar{\chi}_{\langle}}
\newcommand{\cg}{\chi_{\rangle}}
\newcommand{\bcg}{\bar{\chi}_{\rangle}}
\newcommand{\cross}[1]{#1\!\!\!/}
\newcommand{\cald}[1]{{\cal{D}} [#1]}
\newcommand{\intpo}{\int_{-\infty}^{+\infty}\frac{dp^0}{2\pi i}}
\newcommand{\ints}{\int_{-\Lambda}^{\Lambda}ds}
\newcommand{\intv}{\int\frac{d\hat{{\bf v}}}{4\pi}}
\newcommand{\intpos}{\int\frac{dp^0}{2\pi i}\int_{-\Lambda}^\Lambda ds}
\newcommand{\piak}[3]{{\cal{P}}_{#1}^{#2}(#3)}
\newcommand{\piakiak}{{\cal{P}}_{I}^{a}(k)}
\newcommand{\tilv}{\tilde{V}}
\newcommand{\mbyp}{\,\frac{\mu^2}{\pi^2}\,}
\def \Tr {{\rm Tr}}

\def\slr#1{\setbox0=\hbox{$#1$}           % set a box for #1
   \dimen0=\wd0                                 % and get its size
   \setbox1=\hbox{/} \dimen1=\wd1               % get size of /
   \ifdim\dimen0>\dimen1                        % #1 is bigger
      \rlap{\hbox to \dimen0{\hfil/\hfil}}      % so center / in box
      #1                                        % and print #1
   \else                                        % / is bigger
      \rlap{\hbox to \dimen1{\hfil$#1$\hfil}}   % so center #1
      /                                         % and print /
   \fi}

\def\be{\begin{eqnarray}}
\def\ee{\end{eqnarray}}

\begin{document}

\title{The rigidity of crystalline color superconducting quark matter}
\date{August 7, 2007}
\author{Massimo~Mannarelli}
\email{massimo@lns.mit.edu} \affiliation{Center for Theoretical
Physics, Massachusetts Institute of Technology, Cambridge, MA 02139}
\author{Krishna~Rajagopal}
\email{krishna@lns.mit.edu} \affiliation{Center for Theoretical
Physics, Massachusetts Institute of Technology, Cambridge, MA 02139}
\author{Rishi~Sharma}
\email{sharma@mit.edu} \affiliation{Center for Theoretical Physics,
Massachusetts Institute of Technology, Cambridge, MA 02139}

\preprint{MIT-CTP-3807}

\begin{abstract}
We calculate the shear modulus of crystalline color
superconducting quark matter, showing that this phase
of dense, but not
asymptotically dense, three-flavor quark matter responds
to shear stress like a very rigid solid.
To evaluate the shear modulus, we derive the low energy
effective Lagrangian that describes the phonons that originate from  the
spontaneous breaking of
translation invariance by the spatial modulation of the gap parameter $\Delta$.
These massless bosons describe
space- and time-dependent fluctuations of the crystal structure and
are analogous to the phonons in ordinary crystals.  The coefficients
of the spatial derivative terms of the phonon effective Lagrangian
are related to the elastic moduli of the crystal;
the coefficients that encode the linear response of the crystal to a
shearing stress define the shear modulus.  We analyze
the two particular crystal structures which
are energetically
favored over a wide range of densities,
in each case evaluating the phonon effective action and the shear modulus
up to order $\Delta^2$ in a Ginzburg-Landau expansion, finding shear
moduli which are 20 to 1000 times larger than
those of neutron star crusts.
The crystalline color superconducting phase has long
been known to be a superfluid --- by picking a phase
its order parameter breaks the quark-number $U(1)_B$
symmetry spontaneously.  Our results demonstrate
that this superfluid phase of matter is at the same time a rigid solid.
We close with a rough estimate of the pinning force on the rotational
vortices which would be formed embedded within this rigid superfluid
upon rotation.  Our results raise the possibility that
(some) pulsar glitches could originate within a quark matter core
deep within a neutron star.

\end{abstract}
 \maketitle

\section{Introduction}

Quantum chromodynamics (QCD) predicts that
if one can squeeze nuclear matter to the point
that the number of quarks per unit volume is greater
than that within a nucleon,  the matter that results
cannot be described as made up of nucleons: the individual
nucleons are crushed into quark matter~\cite{Collins}.
It remains an open question whether ``liberated'' quark matter
exists at the core of neutron stars, which have central densities
at most $\sim 10$ times that of ordinary nuclear matter.
One route to answering this question would be
an ab initio calculation of the range of densities
at which the expected first order transition from nuclear
to quark matter occurs in QCD.  No currently available
calculational method promises such a result any time soon, which
is not surprising since locating such a first order transition requires
quantitative control of the pressure of both phases, making this
a much more difficult question than understanding the properties of
either phase on its own.  The more promising route to answering
whether neutron stars feature quark matter cores requires
first understanding the qualitative and semi-quantitative
properties of quark matter and then discerning how its
presence deep within a neutron star
would affect astronomically observed or observable phenomena.

Except during the first few seconds after their birth, the
temperature of neutron stars is expected to be of the order of tens
of keV at most. At such low temperatures, if deconfined quark matter exists
in the core of neutron stars it is likely to be in one of the
possible color superconducting phases, whose critical temperatures
are generically of order tens of MeV~\cite{reviews}.
Furthermore,
since compact star temperatures are so much smaller than these
critical temperatures, for many purposes (including in particular
for a calculation of its rigidity) the quark matter that may be
found within neutron stars is well-approximated as having zero temperature,
as we shall assume throughout.

The essence of color superconductivity is quark pairing, driven by
the BCS mechanism which operates whenever there is an attractive
interaction between fermions at a Fermi surface~\cite{BCS}. The QCD
quark-quark interaction is strong and attractive between quarks that
are antisymmetric in color, so we expect cold dense quark matter to
generically exhibit color superconductivity.
It is by now well established that at asymptotic densities, where
the up, down and strange quarks can be treated on an equal footing
and the disruptive effects of the strange quark mass can be neglected,
quark matter is in the color flavor locked (CFL)
phase~\cite{Alford:1998mk,reviews}. The CFL condensate is
antisymmetric in color and flavor indices and therefore involves
pairing between quarks of different colors and flavors.

To describe quark matter as may exist in the cores of compact stars,
we need to consider quark number chemical potentials $\mu$ of order
$500$~MeV at most (corresponding to $\mu_B=3\mu$ of order $1500$~MeV).
The relevant range of $\mu$ is low enough that the strange quark
mass, $M_s$, lying somewhere between its current mass of order
$100$~MeV and its vacuum constituent mass of order $500$~MeV, cannot
be neglected. Furthermore, bulk matter, if present in neutron stars,
must be in weak equilibrium and must be electrically and color neutral. All
these factors work to separate the Fermi momenta of the quarks by
$\sim M_s^2/\mu$, and thus disfavor the cross-species BCS pairing
which characterizes the CFL phase. If we imagine beginning at
asymptotically high densities and reducing the density, and suppose
that CFL pairing is disrupted by the heaviness of the strange quark
before color superconducting quark matter is superseded by baryonic
matter, the CFL phase must be replaced by some other superconducting
phase of quark matter which has less, and less symmetric, pairing.

Considering  only homogeneous condensates, the next phase down in
density is the gapless CFL (gCFL)
phase~\cite{Alford:2002kj,Alford:2003fq,Alford:2004hz,Alford:2004nf,Ruster:2004eg,Fukushima:2004zq,Alford:2004zr,Abuki:2004zk,Ruster:2005jc}.
In this phase, quarks of all three colors and all three flavors
still form ordinary Cooper pairs, but there are regions of momentum
space in which certain quarks do not succeed in pairing, and these
regions are bounded by momenta at which certain fermionic
quasiparticles are gapless. This variation on BCS pairing --- in
which the same species of fermions that pair feature gapless
quasiparticles --- was first proposed for two flavor quark
matter~\cite{Shovkovy:2003uu} and in an atomic physics
context~\cite{Gubankova:2003uj}.  In all these contexts, however,
the gapless paired state turns out in general to suffer from a
``magnetic instability'': it can lower its energy by the formation
of counter-propagating
currents~\cite{Huang:2004bg,Casalbuoni:2004tb}.

In Refs.~\cite{Casalbuoni:2005zp,Mannarelli:2006fy,Rajagopal:2006ig}
the possibility that the next superconducting phase down in density
has a crystalline structure was explored. (For other possible patterns
of three-flavor and three-color pairing,
see Refs.~\cite{Kryjevski:2005qq,Gerhold:2006dt}. For possibilities in
which only two colors of quarks pair, see Refs.~\cite{Gorbar:2005rx}. For possibilities
in which only quarks of the same flavor pair, see Refs.~\cite{Iwasaki:1994ij}. 
See Ref.~\cite{Rajagopal:2005dg} for
an exhaustive study of less symmetric phases with BCS pairing only, the
simplest example being the 2SC phase~\cite{Alford:1997zt,Rapp:1997zu}, indicating
that none are favorable.) Crystalline color
superconductivity --- the QCD analogue of a form of non-BCS pairing
first considered by Larkin, Ovchinnikov, Fulde and Ferrell
(LOFF)~\cite{LOFF} --- is an attractive candidate phase in the
intermediate density regime
\cite{Alford:2000ze,Bowers:2001ip,Kundu:2001tt,Leibovich:2001xr,Bowers:2002xr,Mannarelli:tot},
because it allows quarks living on split Fermi surfaces to pair with
each other. It does so by allowing Cooper pairs with nonzero total
momentum $2{\bf q}^a$, for some set of ${\bf q}^a$'s whose magnitude is fixed
and is of
order the splitting between Fermi surfaces, and whose directions must be
determined.
In position space, this corresponds to condensates that vary
in space like $\sum_a \exp(2 i {\bf q^a}\cdot {\bf x})$, meaning that
the ${\bf q}^a$'s are the reciprocal vectors which define the crystal
structure of the condensate which is modulated periodically in space
and
therefore spontaneously breaks space translation invariance.
These phases seem to be free from magnetic instability~\cite{Ciminale:2006sm},
which is consistent with the result that many of them have free energies
that are lower than that of the (unstable) gCFL phase for wide ranges
of parameter values~\cite{Rajagopal:2006ig}.

Two crystal structures for three-flavor  crystalline
color superconducting quark matter
have been identified as
having particularly low free energy~\cite{Rajagopal:2006ig}.
In the first, called the 2Cube45z phase, the $\langle ud\rangle$
and $\langle us\rangle$ condensates each form
face-centered cubes in position space. (In momentum space,
each has eight ${\bf q}^a$ vectors pointing toward the corners
of a cube, which is the Bravais lattice
of a face-centered cube.) The $\langle us \rangle$ cube is rotated relative
to the $\langle ud \rangle$ cube by 45 degrees. In the second
favorable crystal structure, called the CubeX
phase, the $\langle ud\rangle$ and
$\langle us \rangle$ condensates each describe pairing with
four different wave vectors ${\bf q}^a$, with the eight ${\bf q}^a$'s
together forming the Bravais lattice of a single face-centered cube.
The lattice spacing for these crystal structures is many tens of fm,
as we shall discuss further in Section II, where we specify
these crystal structures precisely.

These phases were identified upon making two crucial
approximations~\cite{Rajagopal:2006ig}, which we shall
also employ. First, the QCD interaction
between quarks was replaced by a point-like Nambu Jona-Lasinio (NJL)
interaction with the quantum numbers of single gluon exchange.
This is a conservative approach to the study of crystalline
color superconductivity, since a point-like interaction
tends to disfavor crystalline color superconductivity
relative to a forward-scattering dominated interaction like single gluon
exchange~\cite{Leibovich:2001xr}.
Second, the calculation employed a Ginzburg-Landau
expansion in powers of the gap parameter $\Delta$ even though
the values of $\Delta$ found in the calculation are large enough
that the control parameter for this expansion, $\Delta/(M_s^2/8\mu)$,
is about 1/2, meaning that the expansion is pushed to or
beyond the
boundary of its regime of quantitative validity.
Upon making these approximations, one or other of the two
most favorable crystalline color superconducting phases
is favored (i.e. has lower free energy than unpaired quark
matter or any spatially uniform paired phase including the
CFL and gCFL phases) for~\cite{Rajagopal:2006ig}
\begin{equation}
2.9\,\Delta_0 < \frac{M_s^2}{\mu}<10.4\,\Delta_0\ .
\label{crystallineregime}
\end{equation}
Here, $\Delta_0$ is the gap parameter in the CFL phase with $M_s=0$, which
is expected to lie somewhere between 10 and 100 MeV~\cite{reviews}.
The estimate~(\ref{crystallineregime}) is valid upon assuming $\Delta \ll M_s^2/8\mu$,
$\Delta_0 \ll \mu$ and $M_s^2\ll\mu^2$,
which here means $10.4\,\Delta_0<\mu$.
In reality, both $M_s$ and $\Delta_0$ are functions of $\mu$, and the 
regime~(\ref{crystallineregime}) translates into a window in $\mu$.  See 
Ref.~\cite{Ippolito:2007uz}
for an analysis in which both $M_s(\mu)$ and $\Delta_0(\mu)$ are calculated
self-consistently in an NJL model, confirming that the window in $\mu$
corresponding to (\ref{crystallineregime}) is broad indeed.
Even though it may not be quantitatively reliable, the
breadth of the regime~(\ref{crystallineregime}),
corresponding to the remarkable robustness of the two
most favorable crystalline phases which can
be understood on qualitative grounds~\cite{Rajagopal:2006ig},
makes it plausible that
quark matter at densities accessible in neutron
stars, say with $\mu\sim 350-500$~MeV,
will be in a crystalline phase. (Unless $\Delta_0$ lies at the upper
end of its allowed range, in which case quark matter at accessible
densities will be in the CFL phase.)

In this paper, we study the elastic  properties of the   CubeX and
the 2Cube45z crystalline phases of
three-flavor
quark matter. We evaluate the  shear moduli of these two
structures  by computing, in the Ginzburg-Landau approximation, the
effective Lagrangian for the phonon modes which emerge due to the
spontaneous breaking of translation invariance by the crystalline
condensates. (See Ref.~\cite{Mannarelli:tot} for  an analysis of
phonons in 2-flavor crystalline color superconducting phases.)
The
shear modulus is related to the coefficients of the spatial
derivative terms that appear in the phonon effective Lagrangian. We
find that when these two crystalline phases are subject
to shear stresses, they behave like rigid solids with
shear moduli
between 20 and 1000 times
larger than those characteristic of conventional neutron star
crusts.   

We evaluate the shear modulus
to order $\Delta^2$ in the Ginzburg-Landau expansion; higher
order terms are suppressed only by powers of $[\Delta/(M_s^2/8 \mu)]^2$,
which is of order 1/4 in the most favorable crystalline phases, large enough
that significant higher order corrections to the shear modulus
can be expected.   However,
as we shall discuss below all that is required from the
calculation of the shear modulus in making the case
that pulsar glitches could originate in a crystalline quark matter
neutron star core is confidence that the shear modulus is comparable to
or greater than that of the conventional neutron star crust.  Since we
find shear moduli that meet this criterion by 
a factor of 20 to 1000,
there is currently
little  motivation for evaluating higher order corrections.

The possibility that quark matter could occur in a solid
phase has been raised previously by Xu~\cite{Xu:2003xe}.  He 
and his collaborators have explored some
astrophysical consequences of a speculation that the
quarks themselves could somehow be arranged in a crystalline
lattice.  The crystalline color superconducting phase is very
different in character: being a superfluid, the quarks are certainly
not frozen in place.  Instead, what forms a crystalline pattern
is the magnitude of the pairing condensate. Although it was clear 
prior to the present work that in this phase 
translational invariance is broken just as in a crystal,
given that
this phase is at the same time a superfluid it was not clear (at least
to us) whether it was rigid.  Here, we demonstrate
by explicit calculation that this
phase, which as we have discussed is plausibly the 
only form of quark matter
that arises at densities reached within neutron star cores, is
rigid indeed. Its shear modulus is parametrically of order 
$\Delta^2 \mu^2$, which could have been guessed on dimensional
grounds. The shear modulus 
is in no way suppressed relative to this natural scale,
even though the crystalline color superconducting phase is superfluid.

Our paper is organized as follows. In Section \ref{crystalline} we
describe the crystalline condensates which act as a background upon
which the phonons act as small perturbations, fully specifying
the CubeX and 2Cube45z structures.
In Section III, after fully specifying the NJL model
that we shall use we
introduce small displacements of a general diquark condensate which
breaks translational symmetries. We write a general expression for
the effective action describing these displacement fields, by
integrating out the fermions in the system. The final result  for
the phonon effective action is given in Eq.~(\ref{Seff3}). We
relate the coefficients of the terms in the effective action
involving the spatial derivatives of the displacement fields to the
shear modulus in Subsection \ref{linear response generalities} and
then evaluate these coefficients for the CubeX and 2Cube45z crystals in
Subsections \ref{CubeX} and \ref{2Cube45z} respectively. We end with a
discussion of our results and their consequences in Section V.

We have moved two relevant consistency checks to the Appendices
to maintain continuity. In Appendix \ref{mass zero} we show explicitly that the
displacement fields are massless to all orders in the gap parameter,
as they must be by Goldstone's theorem. In Appendix \ref{single pw} we evaluate the
effective action for a simple ``crystalline'' structure involving
just two flavors of quarks and pairing with only a single wave vector
${\bf q}$. In this
case, the calculation can be done without making an expansion in the gap
parameter, $\Delta$. We  find that the results in the limit of small
$\Delta$ are consistent with the Ginzburg-Landau calculation.

What are the observable signatures of the presence
of quark matter in a crystalline color superconducting phase
within the core of a neutron star?
Prior to our work, the first steps toward calculating the
cooling rate for such neutron stars were taken in
Ref.~\cite{Anglani:2006br}.  Because the crystalline
phases leave some quarks at their respective Fermi
surfaces unpaired, it seems likely
that their neutrino emissivity
and heat capacity will be only quantitatively smaller than
those of unpaired quark matter~\cite{Iwamoto:1980eb}, not parametrically suppressed.
This suggests that neutron stars with crystalline quark matter
cores will cool by the direct URCA reactions,
i.e. more rapidly than in standard cooling scenarios~\cite{Page:2004fy}.
However, many other possible core compositions can
open phase space for direct URCA reactions, making
it unlikely that this will lead to a distinctive
phenomenology~\cite{UnlikelyFootnote}.
In contrast, the enormous rigidity that we have identified
makes the crystalline phases of quark matter unique
among all forms of matter proposed as candidates for
the composition of neutron star cores.  This makes it
particularly important to investigate its consequences.
We close this Introduction and close Section V of our paper
with speculations about the consequences of a rigid
quark matter core for pulsar glitch phenomenology,
following the lead of Ref.~\cite{Bowers:2002xr}.

A spinning neutron star observed
as a  pulsar gradually spins down as it loses rotational energy
to electromagnetic radiation.
But, every once in a while the angular velocity
at the crust of the star is observed to increase suddenly in a dramatic event called a 
glitch.
The standard 
explanation~\cite{Anderson:1975,Alpar:1977,Alpar:1984a,Alpar:1984b,AlparLangerSauls:1984,Alpar:1985,Epstein:1992,Link:1993,Jones:1993,Alpar:1993,Alpar:1994,Alpar:1996,Pines:1985}
for their occurrence requires the presence of
a superfluid in some region of the star which also
features a rigid array of spatial inhomogeneities which
can pin the rotational vortices in the rotating superfluid.
In the standard explanation of pulsar glitches, these
conditions are met in the inner crust of a neutron star
which features a crystalline array of positively
charged nuclei bathed in a  neutron superfluid (and a neutralizing
fluid of electrons).  The angular momentum of the rotating
superfluid is proportional to the density of vortices, meaning
that as the star as a whole slows the vortices ``want'' to move apart.
As they are pinned to a rigid structure, they cannot.  Hence,
 this superfluid component of the star is spinning faster
than the rest of the star.  After sufficient time, when the ``tension''
in the vortices in some region reaches a critical point, there
is a sudden ``avalanche'' in which vortices unpin and rearrange,
reducing the angular momentum of the superfluid.  The rest
of the star, including in particular the surface whose angular
velocity is observed, speeds up.   We see that this
mechanism requires superfluidity coexisting with a
structure that
is rigid enough that it does not easily deform
when vortices pinned to it seek to move, and
an adequate pinning force which pins vortices
to the rigid structure in the first  place.

The crystalline color superconducting phases
are superfluids: their condensates all spontaneously break
the $U(1)_B$ symmetry corresponding to quark number.
We shall always write the condensates as real.  This choice
of overall phase breaks $U(1)_B$, and 
gradients in this phase correspond to supercurrents.
And yet, as we show at length in this paper,
crystalline color superconductors
are rigid solids with large shear moduli. Supersolids 
are another example of rigid 
superfluids~\cite{Andreev:1969,Chan:2004,Son:2005ak}, 
and on this basis
alone one might consider referring to crystalline color superconducting
phases as supersolid. However, supersolids are rigid by virtue of the
presence of a conventional lattice of atoms whereas in a crystalline
color superconductor individual quarks are not fixed in place in a rigid 
structure --- it is the crystalline pattern of the diquark condensate
that is rigid, exhibiting a nonzero shear modulus.  This makes crystalline
color superconductors sufficiently different from supersolids that we
shall refrain from borrowing the term.  

Although our focus in this
paper will be on the rigidity of crystalline color superconductors,
it is worth pausing to consider how a condensate that vanishes
at nodal planes can nevertheless carry a supercurrent.  In an ordinary
superfluid with spatially homogeneous condensate $\phi$, the ground
state (with no current) can be taken to have $\phi$ real and the
excited state with $\phi=\rho\exp i\theta(\rr)$ carries a current
proportional to $\rho^2 \nabla \theta$.  If we now consider
the simple example of a crystalline color superconductor with
a diquark condensate of the form 
\begin{equation}
\Delta(\rr)=\Delta\cos(2qz)\label{coscondensate}\ ,
\end{equation} 
this can carry current in the $x$- and
$y$-directions in the usual fashion but how can a current in the $z$-direction
flow through the nodal planes at which $|\Delta(\rr)|$ vanishes?  The answer
is that in adddition
to making the condensate complex, with a spatially varying
phase, the presence of a current must deform the magnitude of the
condensate slightly
so as to make it nonvanishing at the location of the (former) nodal planes.
A deformation that has all the appropriate properties is
\begin{equation}
\Delta(\rr) = \Delta(\cos(2qz)+i\epsilon\sin(2qz))\; .\label{zcurrent}
\end{equation}
Note that if we write this condensate as 
$\Delta(\rr)=\rho(z) \exp i \theta(z)$ it satisfies
$\rho^2 d\theta/dz = 2 \epsilon q \Delta^2$, meaning that
the current is $z$-independent as required by current conservation.)  
The deformation of the condensate
is small for small $\epsilon$ as is the current, which is proportional
to $\epsilon$. At the location
of the former nodal planes, the magnitude of the condensate is
also proportional to $\epsilon$.  We shall leave the determination
of whether (\ref{zcurrent}) is the lowest energy deformation of
the condensate (\ref{coscondensate}) with current $2 \epsilon q \Delta^2$ 
to future work; the correct current carrying state must be similar,
even if it does not turn out to be given precisely by (\ref{zcurrent}).
We shall also defer the extension of this analysis to the
realistic, and more complicated, CubeX and 2Cube45z crystal
structures. And, finally, we shall leave the analysis of
currents flowing in circles in crystalline condensates --- i.e. 
the construction of 
vortices --- to future work.

%When a 
%sample of a crystalline color superconducting phase is rotated 
%it may respond by creating superfluid vortices that 
%correspond to configurations 
%where the current carried by the condensates 
%has non-zero circulation along a closed contour.
%Furthermore, it is reasonable to expect that
%these superfluid vortices will have lower free energy if
%they are centered along intersections of nodal planes of
%the crystal structure, i.e. along lines along which the
%condensate already vanishes even in the absence of a rotational
%vortex. 

It is reasonable to expect that the superfluid vortices
that will result when crystalline color superconducting phases
are rotated will have lower free energy if they are centered
along intersections of the nodal planes of the underlying
crystal structure, i.e. along lines along which the condensate
already vanishes even in the absence of a rotational vortex.
By
virtue of being simultaneously superfluids
and rigid solids, then, the crystalline
phases of quark matter provide all the necessary conditions
to be the locus in which (some) pulsar glitches
originate.  The shear modulus, which we calculate,
is larger that that of a conventional neutron star crust
by a factor of 20 to 1000, meaning that the 
rigidity of crystalline quark
matter within a neutron star core is more than sufficient for glitches
to originate therein.
In Section V we
also provide a crude estimate of the pinning force on vortices
within crystalline color superconducting quark matter
and find
that it is comparable to the corresponding value for neutron vortices
within
a neutron star crust. Together, our calculation of the shear modulus
and estimate of the pinning force make the
core glitch scenario worthy of quantitative investigation.
The central questions that remain to be addressed are
the explicit construction of vortices in the crystalline phase
and the calculation of their pinning force, as well
as the timescale over which sudden changes in the
angular momentum of the core are communicated
to the (observed) surface, presumably either via
the common electron fluid or via magnetic
stresses.

\section{Three flavor crystalline color superconductivity \label{crystalline}}

\subsection{Neutral unpaired quark matter \label{neutral unpaired}}

We shall consider quark matter containing massless $u$ and $d$
quarks and $s$ quarks with an effective mass $M_s$. The Lagrangian
density describing this system in the absence of interactions is
given by
\begin{equation}
{\cal L}_0=\bar{\psi}_{i\alpha}\,\left(i\,\de\!\!\!
/^{\,\,\alpha\beta}_{\,\,ij} -M_{ij}^{\alpha\beta}+
\mu^{\alpha\beta}_{ij} \,\gamma_0\right)\,\psi_{\beta j}
\label{lagr1}\ \, ,
\end{equation}
where $i,j=1,2,3$ are flavor indices and $\alpha,\beta=1,2,3$ are
color indices and we  have suppressed the Dirac indices. The mass
matrix is given by $M_{ij}^{\alpha\beta} =\delta^{\alpha\beta}\,{\rm
diag}(0,0,M_s)_{ij} $, whereas
$\de^{\alpha\beta}_{ij}=\delta^{\alpha\beta}\delta_{ij}\partial$ and
the quark chemical potential matrix is  given by
\begin{equation}\mu^{\alpha\beta}_{ij}=(\mu\delta_{ij}-\mu_e
Q_{ij})\delta^{\alpha\beta} + \delta_{ij} \left(\mu_3
T_3^{\alpha\beta}+\frac{2}{\sqrt 3}\mu_8 T_8^{\alpha\beta}\right) \,
, \label{mu}
\end{equation} with  $Q_{ij} = {\rm
diag}(2/3,-1/3,-1/3)_{ij} $ the quark electric-charge matrix and
$T_3$ and $T_8$ the Gell-Mann matrices in color space. $\mu$ is the
quark number chemical potential and if quark matter exists in
neutron stars, we expect $\mu$  to be in the range $350-500$ MeV.

In QCD, $\mu_e$, $\mu_3$ and $\mu_8$ are the zeroth components of
electromagnetic and color gauge fields, and the gauge field dynamics
ensure that they take on values such that the matter is
neutral~\cite{Alford:2002kj,Gerhold:2003js}, satisfying
\begin{equation}
\label{neutrality}
\frac{\partial \Omega}{\partial\mu_e} =
\frac{\partial \Omega}{\partial\mu_3} =
\frac{\partial \Omega}{\partial\mu_8} = 0\, ,
\end{equation}
with $\Omega$ the free energy density of the system.
In the present paper we shall employ an NJL model
with four-fermion interactions and no gauge fields. We introduce
$\mu_e$, $\mu_3$ and $\mu_8$ by hand, and choose them to satisfy
the neutrality constraints (\ref{neutrality}).  The assumption of weak equilibrium
is built into the calculation via the fact that for a given color $\alpha$
the chemical potential of the $d$ quark is equal the chemical potential of the
$s$ quark which is equal to the sum of the chemical potential of the $u$ quark
and $\mu_e$.

Let us consider now the effect of a nonvanishing  strange quark
mass  $M_s$. Suppose we start by setting the gauge chemical potentials
to  zero. Then, in weak equilibrium a nonzero $M_s$ implies
that there
are fewer $s$ quarks in the system than $u$ and $d$ quarks, and
hence the system is positively charged. To restore electrical
neutrality, a positive $\mu_e$ is required which tends to reduce the
number of up quarks relative to the number of down and strange
quarks. In the absence of pairing, color neutrality is obtained with
$\mu_3=\mu_8=0.$

To lowest order in $M_s^2/\mu^2$, the effect of a nonzero strange
quark mass can be taken into account by treating the strange quark
as massless, but with a chemical potential that is lowered by
$M_s^2/(2\mu)$ from $\mu+(\mu_e/3)$. Indeed we can expand the Fermi
momentum of the strange quark as
%\begin{widetext}
\begin{equation}
\begin{split}
p_F^s=\sqrt{\left(\mu+\frac{\mu_e}{3}\right)^2-M_s^2}\approx
&\left(\mu+\frac{\mu_e}{3}\right)-\frac{M_s^2}{2\mu}\\
&+{\cal{O}}\left(\frac{M_s^4}{\mu^3}\right)\, \label{pF1},
\end{split}
\end{equation}
%\end{widetext}
and  to this order electric neutrality requires that
$\mu_e=\frac{M_s^2}{4\mu}$, yielding
\begin{eqnarray}
p_F^d &=& \mu+\frac{M_s^2}{12\mu}=p_F^u+\frac{M_s^2}{4\mu}\nonumber\\
p_F^u &=& \mu-\frac{M_s^2}{6\mu}\nonumber\\
p_F^s &=& \mu-\frac{5 M_s^2}{12 \mu} =p_F^u-\frac{M_s^2}{4\mu}\nonumber\\
p_F^e &=& \frac{M_s^2}{4\mu}\, .\label{pF2}
\end{eqnarray}
Now  we need no longer to be careful about the distinction between
$p_F$'s and $\mu$'s, as we can simply think of the three flavors of
quarks as if they have chemical potentials
\begin{eqnarray}
\mu_d &=& \mu_u + 2 \,\delta\mu_3 \nonumber\\
\mu_u &=&p_F^u \nonumber\\
\mu_s &=& \mu_u - 2 \,\delta\mu_2\,, \label{pF3}
\end{eqnarray}
with
\begin{equation}
\delta\mu_3 = \delta\mu_2 = \frac{M_s^2}{8\mu}\equiv \delta\mu
\label{pF4}\, ,
\end{equation}
and we can write the chemical potential matrix as
\begin{equation}
\mu_{ij}^{\alpha\beta} = \delta^{\alpha\beta}\otimes {\rm
diag}\left(\mu_u,\mu_d,\mu_s\right)_{ij}\label{mu matrix}\,.
\end{equation}
The factor $2$ in front of the $\delta\mu$'s in Eq.~(\ref{pF3}) is
taken to be consistent with the notation used in the analysis of
crystalline superconductivity in a two flavor
model~\cite{Alford:2000ze}, in which the two Fermi momenta were
taken to be $\mu\pm\delta\mu$ meaning that they were separated by
$2\,\delta\mu$. In the three flavor case we have to contend with three
combinations of pairs of Fermi surfaces and we define $2\,\delta\mu_I
= |\mu_j - \mu_k|$ where $j$ and $k$ are flavor indices different
from each other and from $I$. The subscripts on the $\delta\mu$'s in
Eq.~(\ref{pF3}) are consistent with this definition.
The most important driver of qualitative  changes in the
physics of quark matter as a function of decreasing $\mu$
is the increase in $\delta\mu=M_s^2/(8\mu)$, the parameter which
governs the splitting between the three Fermi momenta.

Finally, we note that the equality of $\delta\mu_2$ and $\delta\mu_3$ is only
valid to leading order in $M_s^2/\mu^2$. At the next order,
$\mu_e=M_s^2/(4\mu)-M_s^4/(48\mu^3)$ and hence $\delta\mu_3=\mu_e/2$ while
$\delta\mu_2=\delta\mu_3+M_s^4/(16\mu^3)$. The
fact that $\delta\mu_3$ and $\delta\mu_2$
are close to, but not exactly, equal will be used in Appendix \ref{mass
zero} in the explicit demonstration
that the phonons are massless. The consequences
of the fact
that the splitting between the $u$ and $s$ Fermi surfaces is slightly larger
than the splitting between the $u$ and $d$ Fermi surfaces were explored in
Ref.~\cite{Casalbuoni:2006zs}.

\subsection{The crystalline condensate \label{crystalline condensate}}

The description in Section \ref{neutral unpaired} is valid in the
absence of pairing between quarks. The
interaction between quarks  is attractive between two
quarks that are antisymmetric in color. This induces the formation
of Cooper pairs~\cite{BCS}, and the ground state develops a nonzero
diquark condensate which is predominantly antisymmetric in color
indices. It is free-energetically favorable to have condensates that
are antisymmetric in Dirac
indices~\cite{reviews,Alford:1997zt,Alford:1998mk,Alford:2002kj}
and consequently antisymmetric in flavor also. This implies that
only quarks of different flavor form Cooper pairs.

The fact that quarks that like to pair have Fermi surfaces that are split in the
absence of pairing motivates the possibility that quark matter may exist as a
crystalline superconductor in at least some part of the parameter
space~\cite{LOFF,Alford:2000ze,Bowers:2001ip}. This possibility was explored
for three-flavor quark matter in Ref.~\cite{Rajagopal:2006ig} (and for
simpler condensates
in~\cite{Casalbuoni:2005zp,Mannarelli:2006fy}) where
condensates of the form
\begin{equation}
\langle\psi_{i\alpha}(x)  \psi_{j\beta}(x)\rangle \propto C\gamma_5
 \sum_{I=1}^3\
 \Delta_I  \coleps\flaeps \
\sum_{\q{I}{a}\in\setq{I}{}} e^{2i\q{I}{a}\cdot\rr}
 \label{condensate} 
\end{equation}
were analyzed.
As promised, the expression in Eq.~(\ref{condensate}) is
antisymmetric in color, flavor and Dirac indices. For a given $I$
(which runs from $1$ to $3$), $\Delta_I$ represents the strength of
the pairing between quarks whose color is not $I$ and whose flavor
is not $I$.  The periodic modulation in space of the $I$'th
condensate is defined by a
set of momentum
vectors $\setq{I}{}$.  For example, pairing between $u$ and $d$ quarks
occurs for $ud$-Cooper pairs with any momentum in the set $\setq{3}{}$.
%describing pairing between quarks whose color and flavor are both
%different from $I$.
To shorten notation, we will often refer to the
$\Delta_I$ collectively as $\Delta$ and will henceforth write
$\q{I}{a}\in\setq{I}{}$ simply as $\q{I}{a}$.

The free energy,
$\Omega$, and the gap parameters, $\Delta$, have
been evaluated for many crystalline phases with
condensates of the form (\ref{condensate}) for varied crystal
structures (i.e. varied $\setq{I}{}$)
within the weak coupling and Ginzburg-Landau approximations
$\Delta\ll\delta\mu\ll\mu$~\cite{Rajagopal:2006ig}.

In BCS-paired phases (like the CFL phase for example) quarks
that pair have equal Fermi momenta even if in the absence of
pairing their Fermi momenta would be split. This means that
in such phases, the chemical potentials $\mu_e$, $\mu_3$ and $\mu_8$
take on different values than in the absence of pairing, with the differences
being of order $M_s^2/\mu$.  This rearrangement of Fermi surfaces
exacts a free energy cost.  The reason why crystalline phases can
be more favorable is that they need pay no such cost: the quarks pair
without rearranging Fermi surfaces; the gauge chemical potentials are
as they would be in the absence of pairing, up to corrections that
are proportional to $\Delta$  and thus negligible in the Ginzburg-Landau
approximation.  We therefore  set $\mu_e=M_s^2/(4\mu)$ and $\mu_3=\mu_8=0$
throughout, as in Ref.~\cite{Rajagopal:2006ig}.

The free energy of a crystalline superconductor in the
Ginzburg-Landau approximation relative to that for unpaired
quark matter (i.e.~the condensation energy) can be written as
\begin{widetext}
\begin{equation}
\begin{split}
\Omega_{\rm crystalline}(\{\Delta_I\})= &\frac{2\mu^2}{\pi^2}\Biggl[\sum_I P_I
\alpha_I \, \Delta_I^*\Delta_I \\
&+\ha\Biggl(\sum_I \beta_I(\Delta_I^*\Delta_I)^2
+\sum_{I>J} \beta_{IJ}\, \Delta_I^*\Delta_I\Delta_J^*\Delta_J\Biggr)\\
&+\frac{1}{3}\Biggl(\sum_I \gamma_I(\Delta_I^*\Delta_I)^3
+\sum_{I\neq J}\gamma_{IJJ}\,
\Delta_I^*\Delta_I\Delta_J^*\Delta_J\Delta_J^*\Delta_J
+\gamma_{123}\,\Delta_1^*\Delta_1\Delta_2^*\Delta_2\Delta_3^*\Delta_3\Biggr)\Biggr]
+ {\cal{O}}(\Delta^8)
\label{GLexpansion}\;,
\end{split}
\end{equation}
\end{widetext}
where $P_I$ is the number of momentum vectors in $\setq{I}{}$. The
coefficients appearing in Eq.~(\ref{GLexpansion}) have been calculated for
several crystal structures in an NJL model~\cite{Rajagopal:2006ig},
building
upon previous analysis of the two-flavor case~\cite{Bowers:2002xr}.
The quadratic
coefficient, $\alpha_I$ depends only upon the magnitude of the
momentum vectors in the set $\setq{I}{}$ and the Fermi surface
splitting, $\dm{I}$, and is given by
\begin{widetext}
\begin{equation}
\alpha_I=\alpha(\qia,\dmi)=\Bigl[-1-\ha\log\left(\frac{2^{\frac{2}{3}}\Delta_0^2}{(\qia)^2-\dm{I}^2}\right)
+\frac{\dm{I}}{2|\q{I}{a}|}\log\left(\frac{|\q{I}{a}|+\dm{I}}{|\q{I}{a}|-\dm{I}}\right)\Bigr]\label{alpha}\;.
\end{equation}
\end{widetext}
To find the most favorable magnitude of the vectors $\q{I}{a}$ we
need to minimize the free energy with respect to $|\q{I}{a}|$ which
at leading order is equivalent to minimizing
$\alpha_I$~\cite{Bowers:2002xr}. This fixes the ratio between the
magnitude of the momentum vectors $\q{I}{a}$ and $\dmi$ to be the
same for all the vectors,
\begin{equation}
q_I=|\q{I}{a}|=\eta \,
\dm{I}\label{eta}\;\hspace{.5cm}\forall\;I,\;a\,,
\end{equation}
where $\eta$ satisfies the relation,
\begin{equation}
1+\frac{1}{2\eta}\log\left(\frac{\eta-1}{\eta+1}\right)=0\label{eta
condition}\;,
\end{equation}
and is given numerically by $\eta=1.1997$.

The higher order coefficients in Eq.~(\ref{GLexpansion}) depend upon
the relative directions $\hatq{I}{}$ of the vectors, and the task of
finding favorable structures  involves choosing
polyhedra whose vertices are given by $\hatq{I}{}$,
calculating $\Omega_{\rm crystalline}$ for each choice, and finding those which have
the lowest $\Omega_{\rm crystalline}$.  For a given choice of crystal structure, the
quartic coefficients  in (\ref{GLexpansion}) (i.e. the $\beta$'s)
are dimensionless numbers times $\delta\mu^2$ and the
sextic coefficients (i.e. the $\gamma$'s) are dimensionless
numbers times $\delta\mu^4$.  This makes it clear that
the control parameter for the Ginzburg-Landau approximation
is $\Delta^2/\delta\mu^2=\Delta^2/(M_s^2/8\mu)^2$, which is
of order 1/4 for the most favorable crystal structures~\cite{Rajagopal:2006ig}.
Because we shall
only evaluate the phonon effective action and the shear modulus
to order $\Delta^2$, we shall not need the values
of the $\beta$'s and $\gamma$'s
calculated in Ref.~\cite{Rajagopal:2006ig}; their role in the present calculation
is only indirect, in the sense that they determine that the CubeX and 2Cube45z
structures are most favorable.

The free energy (\ref{GLexpansion}) simplifies upon taking
$\Delta_1=0$.  $\Delta_1$ can be neglected
because it describes
 pairing between $s$ and $d$ quarks, whose Fermi momenta are
 separated by $\dm{1}$ which is twice $\dm{2}$ and $\dm{3}$.

The analysis of Ref.~\cite{Rajagopal:2006ig} yielded qualitative
arguments that  two crystal
structures called CubeX  and
2Cube45z should be most favorable, and explicit calculation
of their free energy to order $\Delta^6$ in the Ginzburg-Landau expansion
showed that they
have comparable free energy to each other and
have a lower (i.e. more favorable) free energy than all the other crystal structures
considered. Furthermore, these two crystalline phases
are favored with respect to
unpaired quark matter and spatially uniform paired phases
including the CFL and gCFL phases over the wide range
of parameters (\ref{crystallineregime}).
For example, taking
$\Delta_0=25$MeV and considering a fixed value of the strange quark
mass $M_s=250$MeV, it turns out that one or other of
the CubeX and 2Cube45z phases is favored
for $\mu$ between
$240$~MeV and $847$~MeV, which more than covers the entire
range of densities relevant for neutron star interiors.
We shall now describe the CubeX and 2Cube45z crystal structures.

\subsection{The CubeX and 2Cube45z structures}

\begin{figure*}[t]
\label{bothcubesfigure}
\begin{center}
%\begin{minipage}[c]{0.45\linewidth}
\includegraphics[width=8cm,angle=0]{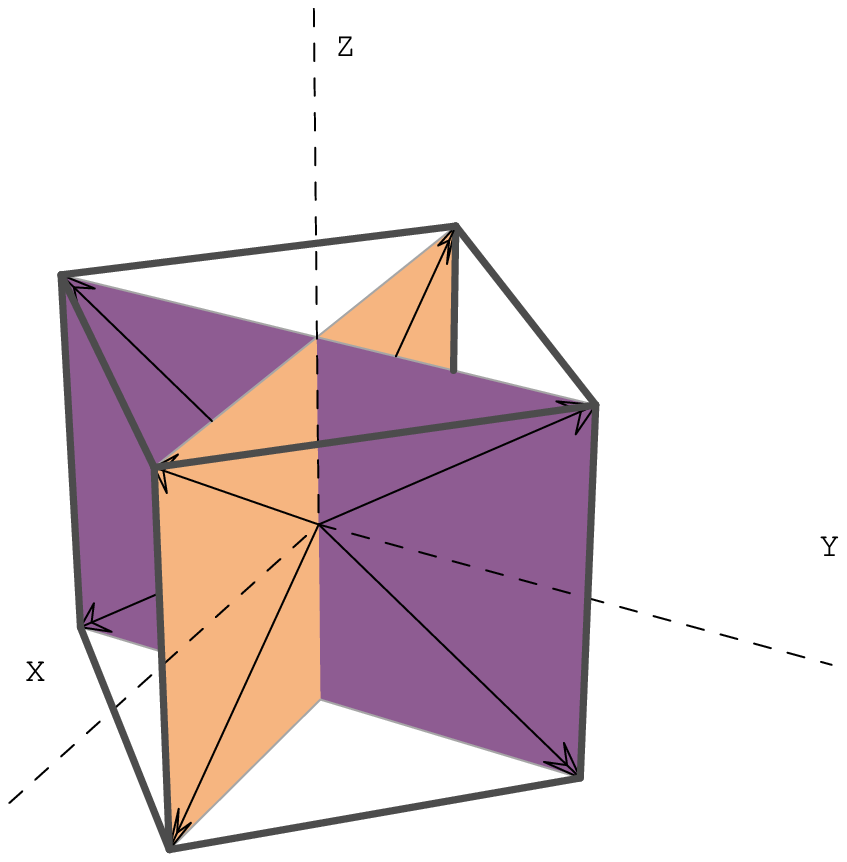}
%\end{minipage}
%\begin{minipage}[c]{0.45\linewidth}
\includegraphics[width=8cm,angle=0]{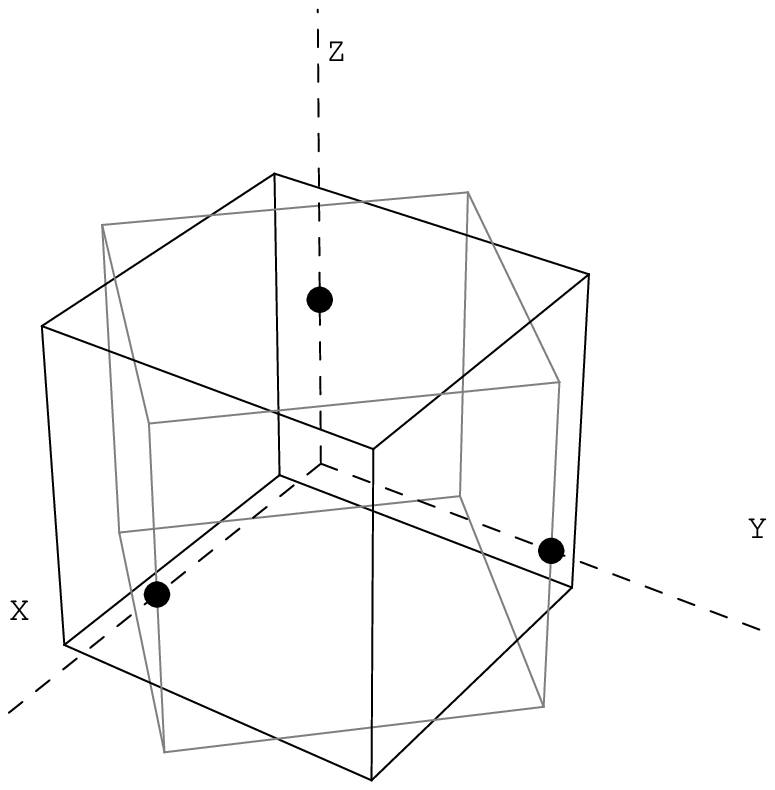}
%\end{minipage}
\caption{(Color online) Left panel: The momentum vectors forming the
CubeX crystal structure.  This structure consists of eight vectors that
belong to two sets $\hatq{2}{}$ and $\hatq{3}{}$ which are shown as
vectors which start from the origin. The four vectors
in $\hatq{2}{}$ are given by ${(1/\sqrt{3})(\pm\sqrt{2},0,\pm 1)}$
and point toward the vertices of the light shaded rectangle (pink
online) that lies in the $x-z$ plane. The four vectors in
$\hatq{3}{}$ are given by ${(1/\sqrt{3})(0,\pm\sqrt{2},\pm 1)}$ and
point toward the vertices of the  dark shaded rectangle (purple
online) that lies in the $y-z$ plane. Taken together the two sets of
vectors point towards the eight vertices of the light gray colored
cube (only the edges are shown as light gray  segments). Right
panel: The end points of the vectors forming the 2Cube45z crystal
structure. This structure consists of sixteen vectors that belong to
two sets $\hatq{2}{}$ and $\hatq{3}{}$. The eight elements of the
set $\hatq{2}{}$  point towards the vertices of the black  cube
(only the edges are shown), and are given by ${(1/\sqrt{3})(\pm
1,\pm 1,\pm 1)}$. The eight elements of the set $\hatq{3}{}$  point
towards the vertices of the light gray cube,  and are given by
$\{(1/\sqrt{3})(\pm \sqrt{2},0,\pm 1)\}\cup\{(1/\sqrt{3})(0,\pm
\sqrt{2},\pm 1)\}$. The three dots denote the points where the axes
meet the light gray cube, to clarify the orientation of the axes.}
\end{center}
\end{figure*}
%\end{figure}
%\end{widetext}

The CubeX crystal structure is specified by two sets of unit vectors,
$\hatq{2}{}$ and $\hatq{3}{}$ depicted in the left panel of
Fig.~1. Taken together, the two sets of vectors point
towards the eight vertices of a cube. The four vectors in
$\hatq{2}{}$ all lie in a plane and point towards the vertices of a
diagonal rectangle of the cube, while the  four vectors $\hatq{3}{}$
form the complementary rectangle. We will use a coordinate system
such that $\hatq{2}{}$ is given by $\{(1/\sqrt{3})(\pm\sqrt{2},0,\pm
1)\}$ (the four combinations of $\pm$ giving the four vectors in
$\hatq{2}{}$) and $\hatq{3}{}$ is given by
$\{(1/\sqrt{3})(0,\pm\sqrt{2},\pm 1)\}$.

The 2Cube45z crystal structure is specified by two sets of unit vectors,
$\hatq{2}{}$ and $\hatq{3}{}$ depicted in the right panel of
Fig.~1. The two sets $\hatq{2}{}$
and $\hatq{3}{}$  {\it{each}} contains eight vectors that
point towards the vertices of a cube. The
cubes specified by $\hatq{2}{}$ and $\hatq{3}{}$ are rotated
relative to each other by an angle $45^\circ$ about one of their
$C_4$ symmetry axis, passing through their common center. We will
orient the coordinate axes such that $\hatq{2}{}$ is given by
$\{(1/\sqrt{3})(\pm 1,\pm 1,\pm 1)\}$ and $\hatq{3}{}$ by
$\{(1/\sqrt{3})(\pm \sqrt{2},0,\pm 1)\}\cup\{(1/\sqrt{3})(0,\pm
\sqrt{2},\pm 1)\}$, which corresponds to a relative rotation by
$45^\circ$ about the $\hat{z}$ axis.

The lattice spacing for the face-centered cubic crystal structure
is~\cite{Bowers:2002xr,Rajagopal:2006ig}
\be
a=\frac{\sqrt{3}\pi}{q}=\frac{4.536}{\delta\mu}=\frac{36.28 \mu}{M_s^2}\ .
\ee
For example, with $M_s^2/\mu=$100, 150, 200 MeV the lattice spacing
is $a=$72, 48, 36 fm.  The spacing between nodal planes is $a/2$.

\section{The phonon effective action \label{phonon}}

In this Section, we present our calculation of the effective action
for the phonons in crystalline color superconducting phases
of quark matter.  In Subsection A we describe the NJL model
which we use. In  Subsections B and C we introduce the phonon
field and integrate the fermions out, yielding a formal expression
for the phonon effective action.  In Subsections D and E we introduce
the Ginzburg-Landau approximation and evaluate the phonon effective
action to order $\Delta^2$.

\subsection{NJL model and mean field approximation}

As described in Ref.~\cite{Rajagopal:2006ig}, we work in an NJL model in
which the quarks interact via a pointlike four-fermion interaction
with the quantum numbers of single gluon exchange, in the mean field
approximation. By this we mean that we add
the local interaction term
%\begin{widetext}
\begin{equation}
{\cal{L}}_{\rm interaction}\ =\ \frac{3}{8}\,G\,
(\bar{\psi}\,\Gamma\,\psi)(\bar{\psi}\,\Gamma\,\psi)
\label{interactionLagrangian}\
\end{equation}
%\end{widetext}
to the Lagrangian density
(\ref{lagr1}), and treat it in the mean field approximation
as we describe below. Here,
we have suppressed the color, flavor and Dirac indices;
the full expression for the vertex factor is $\Gamma^{A\nu}_{\alpha
i,\beta j} = \gamma^\nu (T^A)_{\alpha \beta}\delta_{i j}$, where the
$T^A$ are the color Gell-Mann matrices. To regulate the ultraviolet
behavior of the loop integrals, we introduce a cutoff $\Lambda$
which restricts the momentum integrals to a shell around the Fermi
sphere: $\mu-\Lambda<|\vec{p}|<\mu+\Lambda$.

The free energy $\Omega$ and the gap parameter $\Delta$ for a
crystalline superconductor in the weak coupling and Ginzburg-Landau
approximations depend upon the NJL coupling $G$ and the cutoff
$\Lambda$ only via $\Delta_0$, the gap parameter for the CFL phase
with no Fermi surface splitting, given by
\begin{equation}
\Delta_0=2^{\frac{2}{3}}\Lambda \exp\left[-\frac{\pi^2}{2\mu^2
G}\right]\label{CFL Delta0}\; .
\end{equation}
Model calculations suggest that $\Delta_0$ should lie between
$10\,$MeV and $100\,$MeV~\cite{reviews}.  In the weak-coupling
approximation, $\Delta_0\ll\mu$ and it is possible to
choose $\Lambda$ such that $\Lambda \gg \Delta_0$ while
at the same time $\Lambda \ll \mu$.

We will see that the explicit dependence on $G$ disappears
completely in our calculations and our results depend on the overall
strength of the interactions only via the values of the gap
parameters $\Delta_I$, which in turn depend on $G$ and $\Lambda$
only via the combination $\Delta_0$~\cite{Rajagopal:2006ig}.
It will nevertheless be useful to keep
the cutoff $\Lambda$ in our loop integrals to see explicitly that all the
pieces involved are finite. The final expressions for the
coefficients of the terms that appear in the phonon effective action
are ultraviolet safe, meaning, they are finite and $\Lambda$-independent
as one takes $\Lambda\rightarrow\infty$.    The weak-coupling approximation
$\Delta_0\ll\mu$ is crucial to this result, as this allows one to take $\Lambda\gg \Delta_0$
while keeping $\Lambda\ll\mu$.  Physically, the weak-coupling assumption
means that the pairing is dominated by modes  near the Fermi surfaces
which means that the gaps, condensation energies
and phonon effective action for the crystalline phases are independent of
the cutoff in the NJL model as long as they
are expressed in terms of the
CFL gap $\Delta_0$: if the cutoff is changed with the NJL
coupling adjusted so that $\Delta_0$ stays fixed, the properties
of the crystalline phases also stay fixed.

In the mean field approximation, the interaction Lagrangian
(\ref{interactionLagrangian}) takes on the form
\begin{widetext}
\begin{equation}
{\cal L}_{\rm interaction}= \ha\bar{\psi}\Delta(\rr)\bar{\psi}^T +
\ha\psi^T\bar{\Delta}(\rr)\psi - \frac{3}{8}\,G\, {\rm
tr}\bigl(\Gamma^T\langle\bar{\psi}^T\bar{\psi}\rangle
  \Gamma\bigr)\langle\psi\psi^T\rangle,
\label{meanfieldapprox}
\end{equation}
\end{widetext}
where, $ {\rm tr}$ represents the trace over color, flavor and Dirac
indices, and where $\Delta(x)$ is related to the diquark condensate
by the relations,
%\begin{widetext}
\begin{equation}
\begin{split}
\Delta(x) &= \frac{3}{4}\,G\,\Gamma\langle\psi\psi^T\rangle\Gamma^T\\
%\;{\mbox{, and, }}\;
\bar{\Delta}(x) &= \frac{3}{4}\,G \,
\Gamma^T\langle\bp^T\bp\rangle\Gamma \\
   &=\gamma^0\Delta^{\dagger}(x)\gamma^0 \label{deltaislambdacondensate}\;.
\end{split}
\end{equation}
%\end{widetext}

\subsection{Introduction of the phonon field \label{introduce phonons}}

We now consider the space- and time-dependent vibrations of the
condensate, which will lead us to the effective Lagrangian for
the phonon fields in the presence of a background condensate of the form
(\ref{condensate}). More precisely, we consider the condensate
\begin{equation}
\Delta(\rr)=\Delta_{CF}(\rr)\otimes C\gamma^5\label{spin structure}
\end{equation}
with
\begin{equation}
\Delta_{CF}(\rr)_{\cf} = \sum_{I=1}^3\coleps\flaeps \Delta_I\sum_{\q{I}{a}}
 e^{2i\q{I}{a}\cdot \rr}\label{precisecondensate}\ .
\end{equation}
Sometimes we will write
\begin{equation}
\Delta_{CF}(\rr)= \sum_{I=1}^3\coleps\flaeps \Delta_I(\rr)\label{shorthand}\;,
\end{equation}
with
 \begin{equation}
 \Delta_I(\rr)\equiv\Delta_I\sum_{\q{I}{a}}
 e^{2i\q{I}{a}\cdot \rr}\ .
 \label{cond phon0}
 \end{equation}
Note that $I=1,2,3$ correspond to $\langle ds \rangle$, $\langle us \rangle$
and $\langle ud \rangle$ condensates, respectively.
When we evaluate the phonon effective
Lagrangian  for  the CubeX and 2Cube45z crystals
explicitly in Section IV, we will set $\Delta_1=0$ and $I$ will then run
over $2$ and $3$ only. The condensate (\ref{precisecondensate})
breaks translation invariance spontaneously and we therefore
expect Goldstone bosons corresponding to the broken
symmetries, namely phonons. Phonons are small position and time dependent
displacements of the condensate and, since the three condensates in (\ref{shorthand})
can
oscillate independently, we expect there to be three sets of
displacement fields $\vu_I(x)$. In the presence of the phonons,
then,
\begin{equation}
\Delta_I(\rr)\rightarrow
\Delta_I^u(x)=\Delta_I(\rr-\vu_I(x))\label{cond phon1}\;,
\end{equation}
and we will denote the corresponding quantities appearing on the
left-hand sides of (\ref{shorthand}) and (\ref{spin structure}) as
$\Delta_{CF}^u(x)$ and $\Delta^u(x)$ respectively, i.e.
\begin{equation}
\Delta_{CF}^u(x)=\sum_{I=1}^3\coleps\flaeps \Delta^u_I(x)\label{cond phon2}\;,
\end{equation}
and
\begin{equation}
\Delta^u(x)=\Delta_{CF}^u(x)\otimes C\gamma^5\label{cond phon3}\;.
\end{equation}
(We apologize that we have denoted the displacement fields, and
hence quantities like $\Delta^u$,  by the letter $u$ which in other
contexts, but not here, denotes up quarks.)
In the mean field approximation,
the full Lagrangian, given by the sum of (\ref{lagr1}) and
(\ref{meanfieldapprox}), is quadratic in the fermion fields
and can be written very
simply upon introducing the two component Nambu Gorkov spinor
\begin{equation}
\chi = \left( \begin{array}{c}
\psi    \\
\bp^T
\end{array} \right)\  {\rm and~hence}\
\bar{\chi} = \left( \begin{array}{cc}
\bar{\psi}&\psi^T \end{array}
\right)\label{define chi}\ ,
\end{equation}
in terms of which
\begin{equation}
{\cal L} = \ha \bar{\chi} \left( \begin{array}{cc}
i\cross{\partial}+\cross{\mu} & \Delta^u(x)    \\
\bar{\Delta}^u(x) &   (i\cross{\partial}-\cross{\mu})^T
\end{array} \right) \chi \,+ \frac{1}{16G}
{\rm tr}\bigl((\bar{\Delta}^u)^T\Delta^u\bigr)\;.
\label{fullLagrangian}
\end{equation}
Here, $\cross\mu\equiv \mu\gamma_0$ and $\mu$ is the matrix we have
defined in Eq.~(\ref{mu matrix}). In the next Sections, we shall
also often use the notation $\cross{\mu}_i\equiv \mu_i \gamma_0$,
with $i=1,2,3$ corresponding to $u,d,s$ respectively. The last term
in Eq.~(\ref{fullLagrangian}) comes from the last term in
Eq.~(\ref{meanfieldapprox}), which simplifies to $(1/(16G)){\rm
tr}((\bar{\Delta}^u)^T\Delta^u)$ for condensates given by Eqs.
(\ref{cond phon2}) and (\ref{cond phon3}).

\subsection{Integration over the $\chi$ fields\label{formal integration}}

The spacing between vortices in the
vortex array in a rotating superfluid neutron star is many microns,
and we will be interested in shear stresses exerted over lengths
of order or longer than this length scale.  This means that we need
the effective action for phonon excitations with wavelengths of this order
or longer.  This length scale is many many orders of magnitude longer than
the microscopic length scales that characterize the crystalline phase.
For example, the lattice spacing is many tens of fm.  This means that
we need the effective action for phonons whose wavelength can
be treated as infinite and whose energy can be treated as zero, certainly
many many orders of magnitude smaller than $\Delta$.

The low energy quasiparticles in a crystalline color superconductor
include the
displacement fields $\vu_I(x)$, which are massless because they are
the Goldstone bosons of the broken translational symmetry. In
addition, crystalline superconductors feature gapless fermionic
modes, as we now explain. In the absence of pairing, quarks living at the Fermi surfaces
can be excited without any cost in free energy; pairing in the crystalline
phases yields gaps for quarks living in various ring-shaped bands around
the Fermi surfaces, but leaves gapless fermionic modes at the boundaries
of the pairing regions (loosely speaking, the remainder of the original
Fermi surfaces other than the ring-shaped
bands)~\cite{Alford:2000ze,Bowers:2001ip,Bowers:2002xr,Rajagopal:2006ig}.
The low energy effective theory includes fermions in the vicinity of the
surfaces in momentum space that bound the pairing regions, in addition
to the phonons that are our primary interest in this paper.

To find the low energy
effective action which describes the phonons and the gapless fermionic
excitations we need to integrate out those fermion fields which
have an energy larger than some infrared
cutoff   $\Lambda_{\rm IR}$.  For the application of interest to us,
$\Lambda_{\rm IR}$ should be the energy of phonons with micron
wavelengths.
If we were interested in thermal properties,
$\Lambda_{\rm IR}$ would be of order the temperature $T$.
(Either of these energy scales is $\ll \Delta$, and by the end of this
Subsection we will see that it is safe to set $\Lambda_{\rm IR}=0$.)
In order to formally implement this
procedure, we define \be \psi = \pg + \pll\,~~{\mbox{and hence,
}}~~\bp=\bpg +\bpl\label{mode split psi}\;, \ee where $\pll$ and
$\bpl$ contain modes with energy in $[0,\Lambda_{\rm
IR}]$ and $\pg$ and $\bpg$ those with energy in
$[\Lambda_{\rm IR},\infty]$.
Note that the boundary in momentum space between the $\pg$ and $\pll$
modes will be nontrivial surfaces that follow the boundaries of
the pairing regions, where the gapless fermions are found.
The
corresponding decomposition for the Nambu Gorkov fields is, \be \chi
= \cg + \cll~~\,{\mbox{and hence }}~~\bar{\chi}=\bcg
+\bcl\label{mode split chi}\;, \ee where $\cg$, $\cll$, $\bcg$ and
$\bcl$ are defined analogously to Eq.~(\ref{define chi}). Carrying
out the functional integral over the $\cg$ and $\bcg$ fields will
leave us with a low-energy effective action in terms of the $\vu_I$,
$\cll$ and $\bcl$ fields.

We begin with the path integral expression for the partition
function,
\begin{equation}
Z[\vu,\cll,\bcl]=\int\cald{\cg}\cald{\bcg}e^{i\intspace{x}
{\cal{L}}}\label{define Z}\;,
\end{equation}
where the action of the Lagrangian in Eq.~(\ref{fullLagrangian}) can
be written in terms of the decomposed fields (\ref{mode split chi}),
as follows,
\begin{widetext}
\begin{equation}
\intspace{x}{\cal{L}}=\intspace{x} \Bigl[\frac{1}{16G} {\rm
tr}\bigl((\bar{\Delta}^u)^T\Delta^u\bigr) + \bcg S^{-1}\cg +
\bcl S^{-1}\cll \Bigr]\label{split action}\;,
\end{equation}
\end{widetext}
where the cross terms $\bcg S^{-1}\cll$ and $\bcl S^{-1}\cg$ do not
appear because the integration over $x^0$ imposes energy
conservation, and $\cg$ and $\cll$ have support over different
ranges of energy. The full inverse propagator, $S^{-1}$, in
Eq.~(\ref{split action}) is given by
\begin{equation}
S^{-1}=\left( \begin{array}{cc}
i\cross{\partial}+\cross{\mu} & \Delta^u(x)    \\
\bar{\Delta}^u(x) &   (i\cross{\partial}-\cross{\mu})^T
\end{array} \right)\label{inv prop}\;.
\end{equation}

 Since the Lagrangian is quadratic in the $\chi$ and $\bar{\chi}$ fields, the
standard result for fermionic functional integrals gives
\begin{widetext}
\begin{equation}
i{\cal{S}}[\vu,\cll,\bcl]=\log(Z[\vu,\cll,\bcl])=i\intspace{x} \Bigl[
\bcl S^{-1}\cll + \frac{1}{16G}{\rm tr}
\bigl((\bar{\Delta}^u)^T\Delta^u\bigr)\Bigr] + \ha {\rm Tr}_{{\rm
ng}}\log\left(S^{-1}\right)\label{Z1.5}\;,
\end{equation}
\end{widetext}
where ${\cal{S}}[\vu,\cll,\bcl]$ is the low energy effective action that
we are after, at present still given at a rather formal level.
Here, ${\rm Tr}_{{\rm ng}}$ symbolizes the trace over the Nambu-Gorkov
index, the trace over color, flavor, Dirac indices and the trace
over a set of functions on space-time, with energies lying in
$[-\infty,-\Lambda_{\rm{IR}}]\cup[\Lambda_{\rm{IR}},\infty]$. The
factor $\ha$ appears before ${\rm Tr}_{{\rm ng}}$ because all the
components of $\chi$ and $\bar{\chi}$ are not independent. The
actual independent fields are  $\psi$ and $\bp$. As promised, the
effective action is a function of  the low energy quark fields, which appear in
$\bcl S^{-1}\cll$, and the phonon fields, which appear implicitly via the
dependence of $S^{-1}$ and $\Delta^u$ on $\vu_I$.

We now concentrate on small displacements and hence drop all terms in the
effective action of order $(\vu_{I})^3$ or higher. This is most simply done
by looking at
\begin{widetext}
\begin{equation}
\begin{split}
\Delta^u_{CF}(x)&=\sum_{I=1}^3\coleps\flaeps
\Delta_I\sum_{\qia}e^{2i\qia\cdot(\rr-\vu_{I}(x))},\\
 &= \sum_{I=1}^3\coleps\flaeps
 \Delta_I\sum_{\qia}e^{2i\qia\cdot\rr}
 \left(1-i \fiak{I}{a}{x}-\ha\bigl(\fiak{I}{a}{x}\bigr)^2\right)
 +{\cal{O}}\left(\phi(x)\right)^3\label{cond small phon}\;,
\end{split}
\end{equation}
\end{widetext}
where we have defined
\begin{equation}
2\qia\cdot\vu_{I}(x)=\fiak{I}{a}{x}\label{define phi}\; .
\end{equation}
%where $f_I$ are
%constants of mass dimension $1$, introduced so that the $\phi$'s have mass
%dimension $1$.
We will refer to both the $\vu_I$ fields
and $\phi^a_I$ fields as phonons,
as we can write one in terms of the other.

We now argue that as far as the calculation of the shear modulus is concerned,
we can look only at the part of the effective action that describes the phonons.
 The remainder of  the effective Lagrangian, where the low
 energy quark fields appear,
%we obtain two terms, ${\cal{L}}_{f}$ which refers to the part which
%is quadratic in fermions and does not contain the phonons and
%${\cal{L}}_{f\phi}$ which describes the interaction between the
%quarks and phonon fields.
can be written as
\begin{equation}
\bcl S^{-1}\cll = {\cal{L}}_{f}+{\cal{L}}_{f\phi}
\end{equation}
with
\begin{equation}
{\cal{L}}_{f}= \bcl\left( \begin{array}{cc}
i\cross{\partial}+\cross{\mu} & \Delta(\rr)    \\
\bar{\Delta}(\rr) &   (i\cross{\partial}-\cross{\mu})^T
\end{array} \right)\cll
\end{equation}
and
\begin{equation}
{\cal{L}}_{f\phi}=
\bcl \left( \begin{array}{cc}
0          &  \Delta^u(x)-\Delta(\rr)\\
\bar{\Delta}^u(x)-\bar{\Delta}(\rr)   &  0 \end{array} \right)\cll \label{fermionic part}\;.
\end{equation}
%\bigl(\begin{array}{cc} \pll \\ \bpl^T \end{array} \bigr)
%\bigl(\bpl \;\pll^T\bigr)
We shall see
in Section~\ref{linear response generalities}
that  the shear modulus is related to the coefficient of
\begin{equation}
\frac{\partial\phi}{\partial x^u}\frac{\partial\phi}{\partial x^v}
\label{dphidxdphidx}
\end{equation}
in the Lagrangian, which makes it obvious that
${\cal{L}}_f$ does not contribute.
The coefficient of (\ref{dphidxdphidx})
at the scale $\Lambda_{\rm IR}$ receives
contributions from the $\cg$ and $\bcg$ fermions which have been integrated
out.  The phonons and low energy quarks (at an energy scale
lower than $\Lambda_{\rm IR}$) interact via
the ${\cal L}_{f\phi}$ term in the Lagrangian.
Formally, then, one has to solve consistently
for the phonon propagator and the quark propagator, which are coupled.
However, the effect of the phonon-fermion interactions on the self
consistent calculation of the gap parameter using ${\cal L}_f$ is small,
because
the quark loops come with an additional
factor of $\mu^2$, which is is much larger than $(\Lambda_{\rm IR})^2$, and hence the quark propagator can be considered to be
unaffected by the phonons. The phonon propagator, and hence
the shear modulus, will depend on the phonon-fermion interactions, meaning
that the phonon propagator and consequently
the coefficient of (\ref{dphidxdphidx}) will run as $\Lambda_{\rm IR}$ is reduced.
However, as long as $\Lambda_{\rm IR}$ is much smaller than
$\Delta$, $|{\bf{q}}|$ and $\delta\mu$, the change in the value of the shear
modulus from integrating out more fermions below the scale
$\Lambda_{\rm{IR}}$ will be negligible compared to its value at
$\Lambda_{\rm IR}$. This means that we can take $\Lambda_{\rm IR}=0$,
integrating all of the fermions out from the system and obtaining an effective
action for the phonons alone. This procedure is correct for the
calculation of the shear modulus but would not be correct
for, say, calculating the specific heat of the system, which is dominated
by the gapless fermions not by the phonons. We have
checked numerically that the difference between the shear modulus
calculated with $\Lambda_{\rm IR}=0$ and that with a small but nonzero
$\Lambda_{\rm IR}$ is negligible.

 Finally, therefore, the effective action we are interested in depends only on the
phonon fields, and is given by
\begin{widetext}
\begin{equation}
i{\cal{S}}[\vu]=\log(Z[\vu])=i\intspace{x} \Bigl[
 \frac{1}{16G}{\rm tr}
\bigl((\bar{\Delta}^u)^T\Delta^u\bigr)\Bigr] + \ha {\rm Tr}_{{\rm
ng}}\log\left(S^{-1}\right)\label{Z2}\;,
\end{equation}
\end{widetext}
where now the ${\rm Tr}_{{\rm ng}}$ includes a trace over functions in
space-time containing all energy modes.

 For the single plane wave ``crystal'' structure in which
 only one of the $\Delta_I$ is nonzero and $\{{\bf q}_I\}$
 contains only a single wave vector~\cite{LOFF,Alford:2000ze}, we can
invert the Nambu-Gorkov inverse propagator in the absence of phonons
without expanding in $\Delta$, and can therefore obtain the effective
action for the phonons up to second order in $\phi$,  to {\it all}
orders in $\Delta$. We shall do this exercise in Appendix~\ref{single pw}.
For the realistic crystal structures, CubeX and 2Cube45z,
however, we cannot invert the full inverse propagator  and
we therefore proceed by making a Ginzburg-Landau expansion in
$\Delta$.

\subsection{Ginzburg-Landau expansion\label{ginzburg landau}}

In order to obtain the effective action for the phonon field we
first separate the full inverse propagator, $S^{-1}$, defined in
Eq.~(\ref{inv prop}), into the free part, $S^{-1}_0$ and a part
containing the condensate, $\Sigma$, as follows:
%\begin{equation}
%\begin{split}
${S^{-1}}=S^{-1}_0+\Sigma$,
%\end{equation}
where
\begin{equation}
{S^{-1}_0}=\left( \begin{array}{cc}
i\cross{\partial}+\cross{\mu} & 0    \\
0 &   (i\cross{\partial}-\cross{\mu})^T
\end{array} \right)
\end{equation}
and
\begin{equation}
\Sigma=\left( \begin{array}{cc}
0 & \Delta^u(\rr)    \\
\bar{\Delta}^u(\rr) &   0
\end{array} \right)\ .
\label{split propagator}
\end{equation}
Then, we can expand the term $\log(S^{-1})$ that appears on the
right-hand side of Eq.~(\ref{Z2}) as
\begin{widetext}
\begin{equation}
{\rm Tr}_{{\rm ng}}\bigl(\log({S^{-1}_0}+\Sigma\bigr))={\rm
Tr}_{{\rm ng}}\bigl(\log{S^{-1}_0}\bigr) +{\rm Tr}_{{\rm
ng}}\bigl({S_0}\Sigma\bigr) -\ha {\rm Tr}_{{\rm
ng}}\bigl({S_0}\Sigma\bigr)^2 +...\label{log expansion}\; 
\end{equation}
%\end{widetext}
where ${\rm Tr}_{{\rm ng}}\bigl(\log{S^{-1}_0}\bigr)$ is related to
the free energy of unpaired (Normal) quark matter $\Omega_N$ by
\begin{equation}
%\begin{split}
{\rm Tr}_{{\rm
ng}}\bigl(\log{S^{-1}_0}\bigr)=-i2\Omega_N\intspace{x}
  =-i2VT\Omega_N\;,
%\end{split}
\end{equation}
with $VT$ the space-time volume.
Since
\begin{equation}
\bigl(S_0\Sigma\bigr)=\left( \begin{array}{cc}
0 & (i\cross{\partial}+\cross{\mu})^{-1}\Delta^u(x)    \\
((i\cross{\partial}-\cross{\mu})^T)^{-1}\bar{\Delta}^u(x) &   0
\end{array} \right)\;,
\end{equation}
only even powers of $\bigl(S_0\Sigma\bigr)$ contribute to the trace
over Nambu Gorkov indices and we can write,
%\begin{widetext}
\begin{equation}
{\rm Tr}_{{\rm
ng}}\bigl(\log(S^{-1})\bigr)=-i2\Omega_N(VT)-\sum_{n=1}^\infty
\frac{1}{n} {\rm
Tr}\Bigl((i\cross{\partial}+\cross{\mu})^{-1}\Delta^u(x)
((i\cross{\partial}-\cross{\mu})^T)^{-1}\bar{\Delta}^u(x)\Bigr)^n\label{log
expansion2}\;,
\end{equation}
\end{widetext}
where, the ${\rm Tr}$ on the right hand side is over Dirac, color,
flavor and space-time, and we have used the cyclic property of the
trace to equate the two contributions obtained from the trace over
the Nambu Gorkov index. Finally, substituting (\ref{log expansion2})
back in (\ref{Z2}) and simplifying the Dirac structure of the
operators using the $C\gamma^5$ Dirac structure of the condensate
and the properties of the charge conjugation matrix $C$, namely
$C(\gamma^\mu)^TC^{-1}=-\gamma^\mu$ and $C^2=-1$, we obtain
\begin{widetext}
\begin{equation}
{\cal{S}}[\vu] = -\frac{1}{4G}\intspace{x} {\rm
tr}_{CF}\bigl((\Delta^u_{CF})^\dagger\Delta^u_{CF}\bigr) -
\Omega_N(VT) - \frac{1}{2i}\sum_{n=1}^{\infty} \frac{1}{n} {\rm
Tr}\Bigl((i\cross{\partial}+\cross{\mu})^{-1}\Delta^u_{CF}(x)
(i\cross{\partial}-\cross{\mu})^{-1}\Delta^{u\dagger}_{CF}
(x)\Bigr)^n\label{S1}\;,
\end{equation}
\end{widetext}
where the trace ${\rm tr}_{CF}$ is over color and flavor indices and
where $\Delta^u_{CF}(x)$ depends on $\vu_I(x)$ via Eqs.~(\ref{cond
small phon}) and (\ref{define phi}). Eq.~(\ref{S1}) is
the low energy effective action for the phonons, written
as a Ginzburg-Landau expansion in $\Delta$. We will
calculate the leading contribution to ${\cal{S}}[\vu]$,
namely that proportional to $\Delta^2$.

%\begin{figure}[ht]
\begin{figure*}[t]
\includegraphics[width=13cm,angle=0]{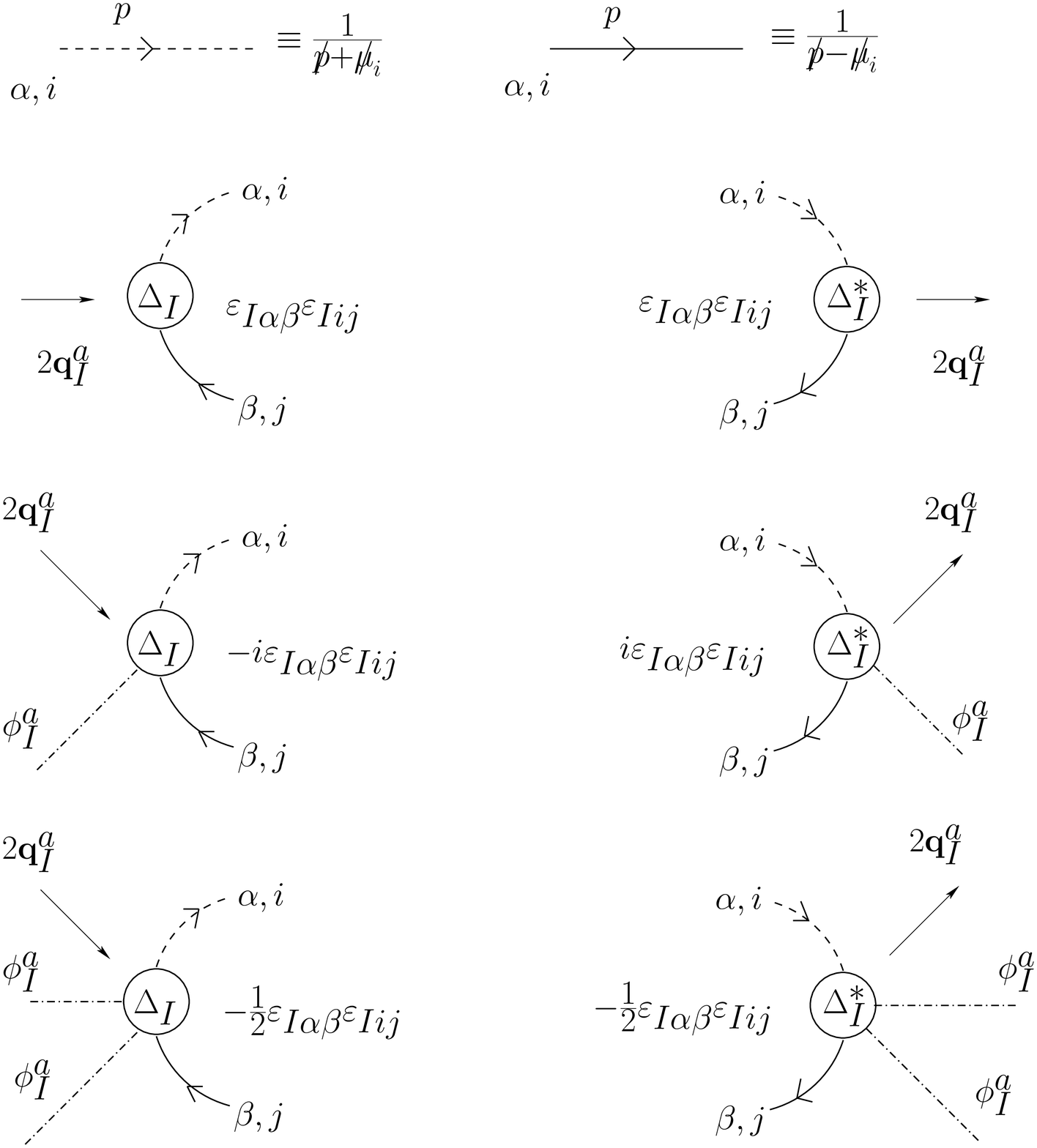}
\caption{Propagators and interaction vertices for the Lagrangian
up to order $\phi^2$. The dashed lines represent propagating
quarks, and the solid lines represent propagating quark holes. The
dot-dash lines represent external phonons. The subscript on $\mu$ is
the index of the flavor which is propagating, and determines the
value of the chemical potential that appears in the propagator. The
$\Delta_I$ vertex comes along with a momentum insertion $2\q{I}{a}$
and a vertex factor $\coleps\flaeps$. Similarly, $\Delta_I^*$ comes
with a momentum insertion $-2\q{I}{a}$ and the same vertex factor.}
\label{feynman rules}
\end{figure*}
%\end{figure}

The first term on the right hand side of Eq.~(\ref{S1}) does not
have any derivatives acting on $\vu_{I}$ and hence can only
contribute to the mass of the phonon, which we know
must be zero by Goldstone's theorem. In Appendix \ref{mass zero}, we show
explicitly that the $\vu_{I}$ dependence in $\intspace{x} {\rm
tr}_{CF}\bigl((\Delta^u_{CF})^\dagger\Delta^u_{CF}\bigr)$ cancels
out, and its value is given simply by
\begin{equation}
\begin{split}
\frac{1}{4G}\intspace{x} {\rm
tr}_{CF}&\bigl((\Delta^u_{CF})^\dagger\Delta^u_{CF}\bigr)\\
&=(VT)\frac{1}{G}\sum_I(\Delta_I\Delta^*_I)P_I\;,
\label{S00}
\end{split}
\end{equation} where $P_I$ is the number of plane waves
in $\setq{I}{}$.

%Moreover in the Appendix \ref{mass zero} we will
%show  that the mass of the phonon fields is zero at any order in
%$\Delta$.

We now proceed to evaluate the third term on the right hand side of
Eq.~(\ref{S1}) diagrammatically. We will expand  the action
${\cal{S}}[\vu]$ in Eq.~(\ref{S1}) in powers of $\phi$ (or
equivalently $\vu_{I}$) up to second order in $\phi$ by using the
Feynman rules described in Fig.~\ref{feynman rules}.

%Term of order phi^0

The lowest order term is independent of $\phi$. The sum of this term
and of the term given in Eq.~(\ref{S00})  has a simple
interpretation. In the absence of phonons, and considering that the
fermionic fields have been integrated out,  the action in
Eq.~(\ref{S1}) turns out to be proportional to the free energy of
the system. More specifically,
\begin{equation}
{\cal{S}}[\vu=0]= -(VT)(\Omega_{\rm crystalline}+\Omega_{N})\;,
\label{Satzero}
\end{equation}
where $\Omega_{\rm crystalline}$ is given as a Ginzburg-Landau series in
$\Delta$~\cite{Rajagopal:2006ig}.  Since (\ref{Satzero}) is independent
of $\phi$ it does not affect the equations of motion of the phonons
and we will simply drop it from our calculation.

%Term of order phi^1
%\begin{widetext}
%\begin{figure}[ht]
\begin{figure*}[t]
\includegraphics[width=13.0cm,angle=0]{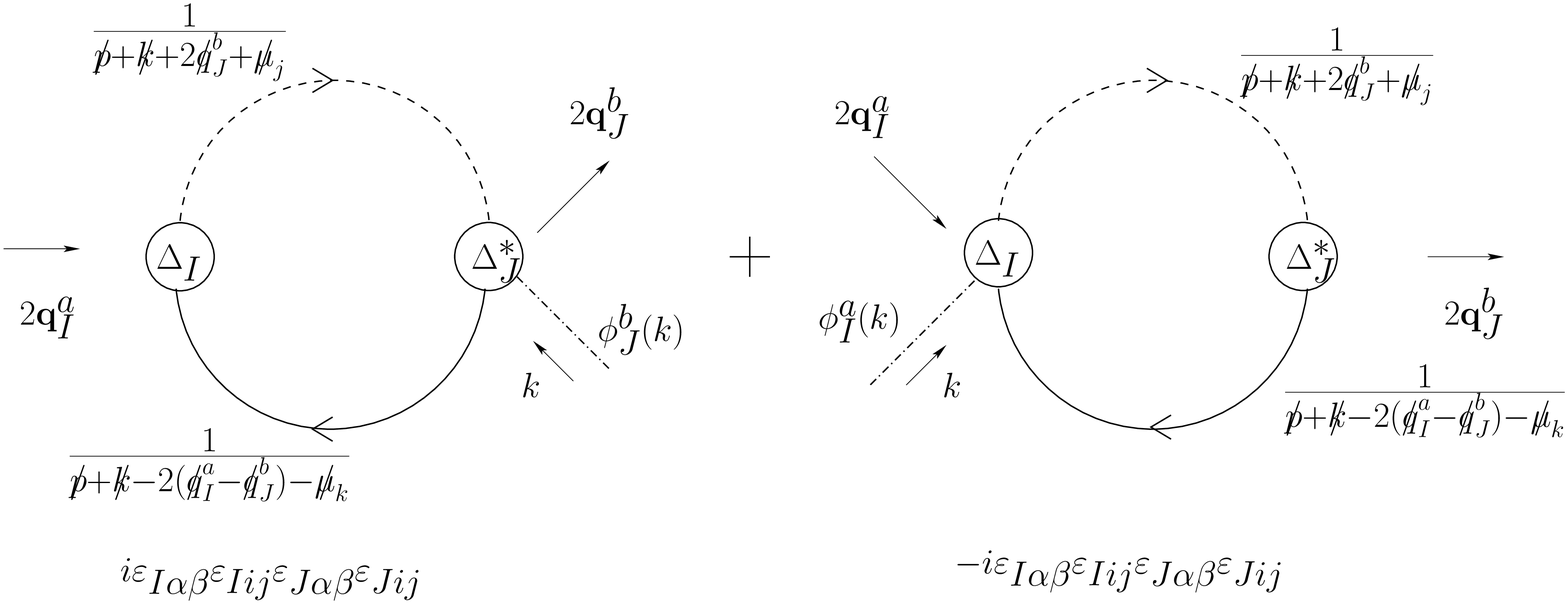}
\caption{Diagrams that contribute to order $\phi \Delta^2$. The
dashed lines represent propagating quarks, and the solid lines
represent propagating quark holes. The interaction vertices have
been defined in Fig.~\ref{feynman rules} and the color-flavor
structure is also indicated. Note that the trace over the
color-flavor epsilon tensors,
$\varepsilon_{I\alpha\beta}\varepsilon_{J\alpha\beta}\varepsilon_{Iij}\varepsilon_{Iij}$
forces $I=J$ and momentum conservation implies $\q{I}{a}=\q{I}{b}$,
as well as $k=0$. The two contributions are then equal in magnitude
and opposite in sign, and hence cancel.} \label{f1diagrams}
\end{figure*}
%\end{figure}
%\end{widetext}

We now consider the term that is linear in $\phi$. We will evaluate the
leading term in the action  proportional to $\Delta^2$, which we
will call ${\cal{S}}^{\phi\Delta^2}$ and which is represented
diagrammatically in Fig.~\ref{f1diagrams}. Both terms shown in
Fig.~\ref{f1diagrams} are proportional to the trace of
$\varepsilon_{I\alpha\beta}\varepsilon_{J\alpha\beta}\varepsilon_{Iij}\varepsilon_{Jij}$,
which is nonzero only if $I=J$ and therefore only terms proportional
to $\Delta_I^*\Delta_I$ are present. We could have anticipated this
result  from the symmetries of the problem. The Lagrangian conserves
particle number for every flavor of quarks, which corresponds to
symmetry under independent global phase rotations of quark fields of
the three flavors, meaning independent phase
rotations of the three  $\Delta_I$. The effective action should be
invariant under these rotations and hence $\Delta_I$ can only occur
in the combination $\Delta_I^*\Delta_I$. (Although the condensate
spontaneously breaks them, the  requirement is that the
Lagrangian has these symmetries.) Then, the sum of the
diagrams in Fig.~\ref{f1diagrams}, which corresponds to the
contribution to the action  linear in the phonon field and
second order in $\Delta$, is given by
\begin{widetext}
\begin{equation}
\begin{split}
{\cal{S}}^{\phi\Delta^2}&=\sum_I\Delta_I^*\Delta_I\sum_{{j\neq
k}\atop{\neq
I}}\sum_{\q{I}{a}\q{I}{b}}\fourier{k}\fourier{p}(2\pi)^4\delta^{(4)}(2\q{I}{a}-2\q{I}{b}+k)\\
&\phantom{+}{\rm tr}\left[
 \frac{\fiak{I}{a}{k}-\fiak{I}{b}{k}}{(\cross{p}+2\cross{q}_I^b+\cross{k}+\cross{\mu}_j)
          (\cross{p}-2(\cross{q}_I^a-\cross{q}_I^{b})+\cross{k}-\cross{\mu}_k)}
 \right]\label{f1value}\;,
\end{split}
\end{equation}
\end{widetext}
where $k$ is the four momentum of the phonon field and the
trace is over Dirac indices. The Dirac delta
on the right-hand side ensures momentum conservation,
\begin{equation}
2\q{I}{a}-2\q{I}{b}+k=0\label{momentum conservation}\;,
\end{equation}
meaning that  the net momentum added to the loop is zero.
 But since we are looking at the low energy effective theory,
we can take $k$  much smaller than the momentum vectors $\q{I}{}$
and therefore Eq.~(\ref{momentum conservation}) can be satisfied
only if $k=0$ and $\q{I}{a}=\q{I}{b}$, which  means that $a=b$.
Using this result,  we find that (\ref{f1value}) vanishes:
\begin{equation}
{\cal {S}}^{\phi\Delta^2}=0\;.
\end{equation}
That is, the term linear in $\phi$ is absent to order $\Delta^2$.
One can similarly argue that it is absent to all orders in $\Delta$.

%Term of order phi^2

%\begin{widetext}
%\begin{figure}[ht]
\begin{figure*}[t]
\includegraphics[width=14.8cm,angle=0]{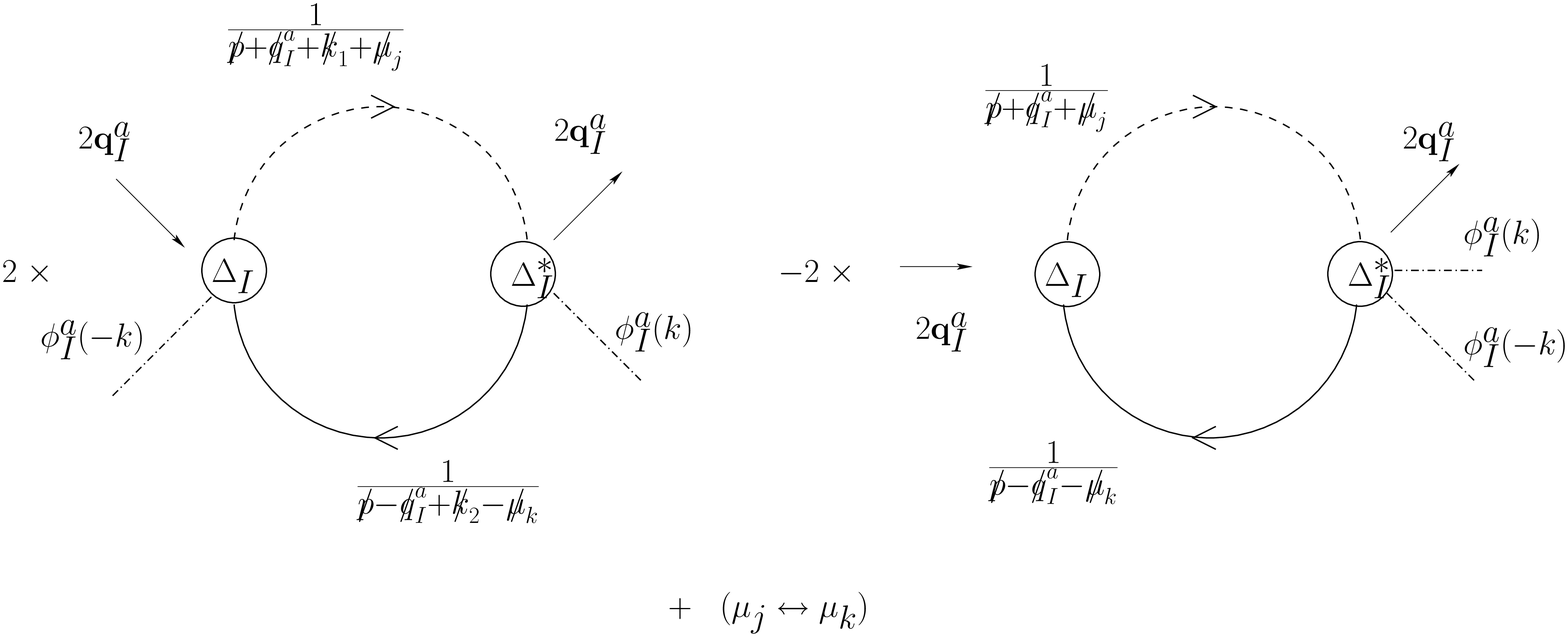}
\caption{Diagrams that contribute to order $\phi^2\Delta^2$. In
drawing the diagrams, we have used the fact that the trace of the
color-flavor  tensor forces $I=J$. We have also used the fact that momentum
conservation requires that the net momenta added by the condensate
and the phonons are separately zero. In the first diagram, momentum
conservation at the $\Delta_I^*$ vertex imposes $k_2-k_1=k$.
%The
%combination $k_2+k_1$ is arbitrary and the value of the diagram
%should be independent of this linear combination.
}
\label{f2diagrams}
\end{figure*}
%\end{figure}
%\end{widetext}

Now we consider the terms of order $\phi^2$, which give the
first nontrivial contribution to the phonon effective
action. We will evaluate these terms to order $\Delta^2$ and we
will call the corresponding contribution to the action
${\cal{S}}^{\phi^2\Delta^2}$. The terms contributing to
${\cal{S}}^{\phi^2\Delta^2}$ arise from the diagrams given in Fig.~\ref{f2diagrams}, and give
\begin{widetext}
\begin{equation}
{\cal{S}}^{\phi^2\Delta^2}=\sum_I
\sum_{\q{I}{a}}\fourier{k}\fiak{I}{a}{k}\fiak{I}{a}{-k}
\Delta_I^*\Delta_I\piak{I}{a}{k}\label{Seff1}\;,
\end{equation}
where $k=k_2-k_1$ is the four momentum of the phonon and
\begin{equation}
\begin{split}
\piak{I}{a}{k} = i\sum_{{j\neq k}\atop{\neq
I}}&\fourier{p}{\rm tr}\Biggl[ \frac{1}{(\cross{p}+\cross{q}_I^a+\cross{k}_1+\cross{\mu}_j)
(\cross{p}-\cross{q}_I^a+\cross{k}_2-\cross{\mu}_k)} \\
-&\frac{1}{(\cross{p}+\cross{q}_I^a+\cross{\mu}_j)
(\cross{p}-\cross{q}_I^a-\cross{\mu}_k)}\Biggr]\label{piak}\;,
\end{split}
\end{equation}
\end{widetext}
where the trace is over Dirac indices.
In the next Subsection we evaluate Eq.~(\ref{piak}). The reader not
interested in the details of our calculation will find in
Eq.~(\ref{Seff2})  the final expression  for the effective action
${\cal{S}}^{\phi^2\Delta^2}$.

\subsection{Evaluation of ${\cal{S}}^{\phi^2\Delta^2}$\label{simplifying piak}}

We turn now to the evaluation of $\piak{I}{a}{k}$ of Eq.~(\ref{piak})
and hence, via (\ref{Seff1}),  the leading nontrivial contribution to the phonon
low energy effective action, ${\cal{S}}^{\phi^2\Delta^2}$.

To begin, if we set $k_1=k_2=0$ in (\ref{piak}), implying that
$k=0$, we see that $\piak{I}{a}{0}=0$.  In this way, we see explicitly that
the phonons are massless to order $\Delta^2$. As mentioned before,
we are interested in the low energy, long wavelength phonons.
We therefore expand $\piak{I}{a}{k}$ in powers
of $k$ and drop terms of order $k^3$ and higher.

We are working in the  limit in which $\dm{}$,
$q=|\q{}{}|=\eta\,\dm{}$ and $\Delta$  are all much smaller than $\mu$.   ($\Delta\ll\mu$
follows from the weak coupling approximation and $\delta\mu\ll\mu$ follows from
requiring
$M_s^2\ll\mu^2$.  The Ginzburg-Landau approximation, which is
the further requirement that $\Delta^2\ll\delta\mu^2$, is not required in
the derivation of the simplifications of Eq.~(\ref{piak}) that follow.)
This
means that we can choose the ultraviolet  cutoff $\Lambda$ such that
$\delta\mu,q,\Delta\ll \Lambda \ll \mu$. Since the largest
contribution to the integrals comes from the region close to the
Fermi surfaces and since $\Lambda\ll\mu$, the integration measure in
Eq.~(\ref{piak}) can be approximated as follows:
\begin{equation}
 i\fourier{p} \approx
\frac{i{\mu}^2}{2\pi^2}\int_{-\infty}^{+\infty}\frac{dp^0}{2\pi}\int_{-\Lambda}^{\Lambda}
ds\int\frac{d\hat{{\bf v}}}{4\pi}\;,
\label{IntegrationMeasure}
\end{equation}
where $\hat{{\bf{v}}}$ is the unit velocity vector in the direction of
the spatial momentum vector, $\hat{{\bf{v}}}={\bf{p}}/|{\bf{p}}|$, and $\int d{\vv}$
represents the integral over solid angle covering the Fermi
surface. The residual momentum $s$ is defined by the relation
$s\equiv |\bf{p}|-\bar{\mu}$, where $\bar{\mu}$ is an energy scale
lying close to the quark Fermi surfaces. In evaluating
$\piak{I}{a}{k}$, we will take $\bar{\mu}$ to be the arithmetic mean
of $\mu_k$ and $\mu_j$, for convenience, but since the two integrals
on the right-hand side of Eq.~(\ref{piak}) go as
$\log(\Lambda)$ for large momenta, choosing any other value for $\bar{\mu}$ close to
the Fermi surfaces changes the value of $\piakiak$ only by
${\cal{O}}(\dm{}/\Lambda)$, which we will ignore. We introduce
the two null vectors, $V^\mu=(1,{\vv})$ and
$\tilde{V}^\mu=(1,-{\vv})$, as is done in the High Density Effective
Theory (HDET)~\cite{Nardulli:2002review}. It is also useful to define four momenta
$l^\mu\equiv(p^0,s\hat{{\bf{v}}})=p^\mu-(0,\bar{\mu}\hat{{\bf{v}}})$, which
can be thought of as residual momenta as measured from the Fermi surface. It is
easy to verify that $V\cdot l = p^0-s$ and $\tilde{V}\cdot l = p^0+s$. In the weak coupling
limit, for a generic four vector $p'$ that is small compared to $\Lambda$, the propagators
in Eq.~(\ref{piak}) simplify as follows:
\begin{widetext}
\begin{equation}
\begin{split}
\frac{1}{\cross{p}+\cross{p}'+\cross{\mu}_j}
&=
 \frac{(p^0+(p')^0+\mu_j)\gamma^0-({\bf  p}+{\bf p'})\cdot{\vec{\gamma}}}
 {(p^0+(p')^0+\mu_j-|{\bf p}+{\bf p'}|)(p^0+(p')^0+\mu_j+|{\bf p}+{\bf p'}|)}\\
&\approx \frac{\bar{\mu}\gamma^0 - {\bf p}\cdot{\vec{\gamma}}}
 {(p^0+(p')^0+\bar{\mu}+(-\bar{\mu}+\mu_j)-|{\bf p}|-{\bf p'}\cdot\hat{{\bf{v}}})(2\bar{\bar{\mu}})}\\
&\approx
 \frac{1}{2}\left(\frac{\gamma^0-\hat{{\bf{v}}}\cdot{\vec{\gamma}}}
 {p^0+(p')^0-s+(\mu_j-\bar{\mu})-{\bf p'}\cdot\hat{{\bf{v}}}}\right)\\
&=
 \frac{1}{2}\left(\frac{V\cdot\gamma}
 {V\cdot(l+p')+(\mu_j-\bar{\mu})}\right) \label{prop1}\;
\end{split}
\end{equation}
and, similarly,
\begin{equation}
\begin{split}
\frac{1}{\cross{p}+\cross{p}'-\cross{\mu}_k}
&=
 \frac{(p^0+(p')^0-\mu_k)\gamma^0-({\bf  p}+{\bf p'})\cdot{\vec{\gamma}}}
 {(p^0+(p')^0-\mu_k-|{\bf p}+{\bf p'}|)(p^0+(p')^0-\mu_k+|{\bf p}+{\bf p'}|)}\\
&\approx
 \frac{1}{2}\left(\frac{\gamma^0+\hat{{\bf{v}}}\cdot{\vec{\gamma}}}
 {p^0+(p')^0+s-(\mu_k-\bar{\mu})+{\bf p'}\cdot\hat{{\bf{v}}}}\right)\\
&=
 \frac{1}{2}\left(\frac{\tilv\cdot\gamma}
 {\tilv\cdot(l+p')-(\mu_k-\bar{\mu})}\right) \label{prop2}\;.
\end{split}
\end{equation}
Upon using these simplifications in
Eq.~(\ref{piak}), we obtain,
\begin{equation}
\begin{split}
\piak{I}{a}{k} &= \frac{\mu^2}{\pi^2}\Bigl[
\intpo\ints\intv\frac{1}{(V\cdot l-\hat{\bf{v}}\cdot\q{I}{a}+\dm{I})
(\tilv\cdot l-\hat{\bf{v}}\cdot\q{I}{a}+\dm{I})} \\
-&\intpo\ints\intv\frac{1}{(V\cdot (l+k_1)-\hat{\bf{v}}\cdot\q{I}{a}+\dm{I})
(\tilv\cdot (l+k_2)-\hat{\bf{v}}\cdot\q{I}{a}+\dm{I})}
\Bigr]+(\dm{I}\rightarrow-\dm{I})\label{piak2}\;.
\end{split}
\end{equation}
\end{widetext}

By making the changes of variables $p^0\rightarrow -p^0$, $s\rightarrow -s$,
$\hat{\bf{v}}\rightarrow -\hat{\bf{v}}$ and $\dm{I}\rightarrow
-\dm{I}$ it is easy to show that
$\piak{I}{a}{-k^0,{\bf{k}}}=\piak{I}{a}{k^0,{\bf{k}}}$. In
addition, it is clear from Eq.~(\ref{Seff1}) that only the part of
$\piakiak$ that is even under $k\rightarrow -k$ contributes to
${\cal S}^{\phi^2\Delta^2}$.
Hence, terms in the expansion of
$\piakiak$ proportional to odd powers of $k$ do not contribute to
the effective action. Furthermore, at second order
in $k$ this implies that there cannot be  terms proportional to
$k_0 {\bf{k}}\phi^2\Delta^2$ in the effective action. (Terms like
$k_0^2 {\bf{k}}^2\phi^2\Delta^2$ can of course appear, but are
higher order in $k$.) This is useful because we can handle the
spatial and time components of $k$ independently, thereby
simplifying the calculation of $\piakiak$.

In order to simplify the calculation we  rewrite $\piakiak$ a little
differently. Multiplying and dividing the integrand appearing in the
second term in Eq.~(\ref{piak2}) (the integrand depending upon $k_1$
and $k_2$) by
\begin{equation}
\begin{split}
&(\tilv\cdot (l+k_1)-\hat{\bf{v}}\cdot
\q{I}{a}+\dm{I})\\
&\times (V\cdot (l+k_2)-\hat{\bf{v}}\cdot \q{I}{a}+\dm{I})
\end{split}
\end{equation}
and collecting the term with numerator\\ $ \bigl(V\cdot k\bigr)
\bigl(\tilv\cdot k\bigr) =
\bigl(V\cdot(k_2-k_1)\bigr)\bigl(\tilv\cdot(k_2-k_1)\bigr)$,\\ after some algebra we obtain
\begin{widetext}
\begin{equation}
\frac{2}{(V\cdot (l+k_1)-\hat{\bf{v}}\cdot \q{I}{a}+\dm{I})
(\tilv\cdot (l+k_2)-\hat{\bf{v}}\cdot \q{I}{a}+\dm{I})}
=-\frac{\bigl(V\cdot k\bigr)\bigl(\tilv\cdot
k\bigr)}{D(l+k_1)D(l+k_2)}+\frac{1}{D(l+k_1)} +\frac{1}{D(l+k_2)}
\label{simplify piak}\;,
\end{equation}
%\end{widetext}
where
\begin{equation}
D(l)\equiv(\tilv\cdot l +\dm{I}-\hat{\bf{v}}\cdot\q{I}{a})(V\cdot l
+\dm{I}-\hat{\bf{v}}\cdot\q{I}{a})\label{define D}\,.
\end{equation}
We can then write $\piakiak$ as $\piakiak=I_0+I_1$ with
%\begin{widetext}
\begin{eqnarray}
I_1&\equiv& \ha\mbyp\intpos\intv \frac{\bigl(V\cdot
k\bigr)\bigl(\tilv\cdot
k\bigr)}{D(l+k_1)D(l+k_2)} + (\dm{I}\rightarrow-\dm{I})\label{I1}\\
I_0&\equiv &\mbyp\intpos\intv\frac{1}{D(l)}
-\ha\mbyp\intpos\intv\frac{1}{D(l+k_1)}\nonumber\\
  &&-\ha\mbyp\intpos\intv\frac{1}{D(l+k_2)}+
  (\dm{I}\rightarrow-\dm{I})\label{I0}\,.
\end{eqnarray}
\end{widetext}
The reason to separate $\piakiak$ as a sum of $I_0$ and $I_1$ will
become clear in a moment, when we argue that $I_0=0$.

We now proceed to evaluate $I_0$ and $I_1$ separately. As we
discussed, we can consider the spatial and the temporal parts of $k$
independently, and we begin by taking $k=(0,\vk)$ with
$\vk={\bf{k}}_2-{\bf{k}}_1$. In this case $I_0$ can be expressed as
a sum of three terms, each proportional (up to a prefactor
$+\mu^2/\pi^2$ or $-\mu^2/(2\pi^2)$)
to an integral of the form
\begin{widetext}
\begin{equation}
\begin{split}
\Pi(\q{I}{a},\dm{I},p')
  &=\intpos\intv \frac{1}{D(l+p')}\\
  &=\intpos\intv \frac{1}{(p^0-s-\hat{\bf{v}}\cdot{\bf{p'}}-\hat{\bf{v}}\cdot\q{I}{a}+\dm{I})
    (p^0+s+\hat{\bf{v}}\cdot{\bf{p'}}-\hat{\bf{v}}\cdot\q{I}{a}+\dm{I})}
  \label{defining Pi}\;,
\end{split}
\end{equation}
%\end{widetext}
where $p'$ is in this case a purely spatial vector,
$p'=(0,{\bf{p'}})$. This integral can be evaluated by following the
steps outlined in Ref.~\cite{Bowers:2002xr}. We first perform a Wick
rotation, $p^0\rightarrow ip^4$, and then do the $p^4$ integration
by the method of residues, followed by the $ds$ and $d\hat{\bf{v}}$
integrals. For ${\bf {p'}}=0$, the integral is calculated in
Ref.~\cite{Bowers:2002xr} and is given by
%\begin{widetext}
\begin{equation}
\Pi(\q{I}{a},\dm{I},0)=\left[-1-\ha\log\left(\frac{\Lambda^2}{(\qia)^2-\dmi^2}\right)
+\frac{\dm{I}}{2|\q{I}{a}|}\log\left(\frac{|\q{I}{a}|+\dm{I}}{|\q{I}{a}|-\dm{I}}\right)\right]
\label{Pi0}\;.
\end{equation}
%\end{widetext}
By making the change of variables $s\rightarrow s-\hat{\bf{v}}\cdot{\bf{p'}} $
in Eq.~(\ref{defining Pi}), we see
that the integrand appearing in the definition of $\Pi(\q{I}{a},\dm{I},p')$
in Eq.~(\ref{defining Pi}), can be written as the integral appearing in
$\Pi(\qia,\dmi,0)$, but with the limits of $s$ integration changed
to
$[-\Lambda-\hat{\bf{v}}\cdot{\bf{p'}},\Lambda-\hat{\bf{v}}\cdot{\bf{p'}}]$.
Since, $\Pi$ goes as $\log(\Lambda)$, however, this change in
limits changes the value of the integrand only by a quantity of
order $\vk/\Lambda$, which we ignore. Thus,
%\begin{widetext}
\begin{equation}
%\begin{split}
I_0 =\mbyp \left[\Pi(\qia,\dmi,0) -
\ha\Pi(\qia,\dmi,0)-\ha\Pi(\qia,\dmi,0)\right] + (\dmi\rightarrow
-\dmi)
  =0\;.
%\end{split}
\end{equation}
%\end{widetext}
The integral $I_1$ can be evaluated using similar steps, namely Wick rotate
$p^0\rightarrow ip^4$, perform the $p^4$ integration by residues and then
do the $ds$ and $d\hat{\bf{v}}$ integrals. The final result is
%\begin{widetext}
\begin{equation}
I_1=\mbyp \left[\Pi(\qia,\dmi,0) -
\ha\Pi(\qia-\ha\vk,\dmi,0)-\ha\Pi(\qia+\ha\vk,\dmi,0)\right]
+ (\dmi\rightarrow -\dmi)\label{I1 result}\;.
\end{equation}
%\end{widetext}
We note that the final result depends only upon $\vk=\vk_2-\vk_1$.
Expanding $I_1$ in $\vk$, we find
%\begin{widetext}
\begin{equation}
\begin{split}
\piakiak&=-\kpa^2\mbyp\left[\frac{1}{4(\q{I}{a2}-\dmi^2)}
  -\frac{1}{2(\qia)^2}\left(1+\frac{\dmi}{2|\qia|}\log\left(\frac{|\qia|-\dmi}{|\qia|+\dmi}\right)\right)\right]\\
  &-\kpe^2\mbyp\left[
  \frac{1}{4(\qia)^2}\left(1+\frac{\dmi}{2|\qia|}\log\left(\frac{|\qia|-\dmi}{|\qia|+\dmi}\right)\right)\right]
  +{\cal{O}}(k^4)
  \label{piak spatial0}\;,
\end{split}
\end{equation}
\end{widetext}
where $\kpa$ is the component of $\vk$ which is parallel to $\qia$,
$\kpa=\unitqia(\vk\cdot\unitqia)$, and $\kpe$ is the component
perpendicular to $\qia$, $\kpe=\vk-\kpa$. In deriving Eq.~(\ref{piak
spatial0}) we did not assume any particular relations between $\qia$
and $\dmi$, but now we choose the value of $|\qia|$ given by
Eq.~(\ref{eta}) that minimizes the free energy. Substituting
Eqs.~(\ref{eta}) and (\ref{eta condition}) into Eq.~(\ref{piak spatial0}) simplifies $\piakiak$
considerably, yielding
\begin{equation}
\piakiak=-\kpa^2\mbyp\left[\frac{1}{4\dmi^2(\eta^2-1)}\right]\label{piak
spatial1}\hspace{0.5cm} {\rm for} \hspace{0.2cm} k=(0,{\bf k})\;,
\end{equation}
where we have dropped the terms of order $k^4$.

The final expression for $\piakiak$ has the following features.
First, the $\kpe$ term has dropped out. This means that for a
single plane wave condensate, phonons that propagate in the
direction orthogonal to the plane wave that forms the condensate
cost no energy up to order $(\kpe)^2\Delta^2\phi^2$. Second, we
note that the coefficient in front of $(\kpa)^2$ is negative.
This
means that
 the crystal structure is stable
with respect to small fluctuations  in the direction of $\qia$. (Recall that
action goes like kinetic energy minus potential energy; since here $k$ is spatial,
we have potential energy only meaning that decreasing the action corresponds
to increasing the energy, hence stability.)
This result
is a direct consequence of the fact that we chose $|\qia|$ to
minimize $\alpha_I$, meaning that
any deviation from the most favorable modulation of the condensate
in the direction of $\qia$ increases the free energy of the system by an
amount of order $\Delta^2$.

We now evaluate $I_0$ and $I_1$ in the case where $k$ is purely
temporal, namely  $k=(k^0,{\bf{0}})$ with $k^0=k^0_2-k^0_1$.
With these $k_1$ and $k_2$, the value of
$I_1$ turns out to be
\begin{widetext}
\begin{equation}
\begin{split}
I_1&=\mbyp\Biggl[
 \ha\log\left(\frac{(\dmi-k^0_1+(k^0/2))^2-(\qia)^2}{(\dmi-k^0_1)^2-(\qia)^2}\right)
+(\dmi-k^0_1+(k^0/2))\log\left(\frac{\dmi-k^0_1+(k^0/2)+|\qia|}{\dmi-k^0_1+(k^0/2)-|\qia|}\right)\\
&-(\dmi-k^0_1)\log\left(\frac{\dmi-k^0_1+|\qia|}{\dmi-k^0_1-|\qia|}\right)
+\ha\log\left(\frac{(\dmi-k^0_2-(k^0/2))^2-(\qia)^2}{(\dmi-k^0_2)^2-(\qia)^2}\right)\\
&+(\dmi-k^0_2-(k^0/2))\log\left(\frac{\dmi-k^0_2-(k^0/2)+|\qia|}{\dmi-k^0_2-(k^0/2)-|\qia|}\right)
-(\dmi-k^0_2)\log\left(\frac{\dmi-k^0_2+|\qia|}{\dmi-k^0_2-|\qia|}\right)\Biggr]+(\dmi\rightarrow
-\dmi)\label{I1 temporal1}\;.
\end{split}
\end{equation}
%\end{widetext}
Although, it appears that Eq.~(\ref{I1 temporal1}) does not depend
solely on $k^0_2-k^0_1$, upon expanding in small $k^0_1$ and
$k^0_2$, we find,
%\begin{widetext}
\begin{equation}
%\begin{split}
I_1=(k^0)^2\mbyp \frac{1}{4((\qia)^2-\dmi^2)} + {\cal{O}}(k^4)
   \approx (k^0)^2\mbyp \frac{1}{4\dmi^2(\eta^2-1)}  \label{I1 temporal2}\;.
%\end{split}
\end{equation}

Turning now  to $I_0$, this quantity is
given by
\begin{equation}
I_0 = \frac{\mu^2}{\pi^2}\left[ \Pi(\q{I}{a},\dm{I},0) -\frac{1}{2}
\Pi(\q{I}{a},\dm{I},k_1) -\frac{1}{2} \Pi(\q{I}{a},\dm{I},k_2)
\right] \label{I0temporal}
\end{equation}
where $k_i=(k_i^0,{\bf{0}})$.  When its third argument is
a purely temporal four-vector, $\Pi$ is given by
%\begin{widetext}
\begin{equation}
\Pi(\q{I}{a},\dm{I},p')
  =\intpos\intv \frac{1}{(p^0+p^{\prime0}-s-\hat{\bf{v}}\cdot\q{I}{a}+\dm{I})
    (p^0+p^{\prime0}+s-\hat{\bf{v}}\cdot\q{I}{a}+\dm{I})}
  \label{defining Pi2}\; ,
\end{equation}
\end{widetext}
where $p'=(p'^0,{\bf{0}})$.
It is apparent from
Eq.~(\ref{defining Pi2}) that by making the change of variables $p^0\rightarrow
p^0+p^{\prime0}$ we obtain
$\Pi(\qia,\dmi,p^{\prime0})=\Pi(\qia,\dmi,0)$, leading us to
conclude from Eq.~(\ref{I0temporal}) that $I_0=0$.
We advise the reader that in
order to obtain the result $I_0=0$ by the approach that we have employed in a
straightforward manner, it is important to shift $p_0$ before Wick rotating.
(Note that if we calculate
$I_0$ using dimensional regularization  
%or by introducing
%a nonzero temperature and then taking the $T \rightarrow 0$ limit, 
we
find $I_0=0$ in agreement with what we obtained by change
of variables.)
Finally, therefore, with $I_0=0$ we obtain
\begin{equation}
\piakiak = (k^0)^2 \mbyp \frac{1}{4\dmi^2(\eta^2-1)} \hspace{0.5cm}
{\rm for} \hspace{0.2cm} k=(k^0,{\bf 0})\;.
\label{piaktemporal}\end{equation} We see that this comes with a
positive sign, as is appropriate for a kinetic energy term.

Substituting the expressions given in  Eqs.~(\ref{piak spatial1})
and (\ref{piaktemporal}) back into the action (\ref{Seff1}), we obtain
\begin{widetext}
\begin{equation}
{\cal{S}}^{\phi^2\Delta^2}=\sum_I\sum_{\qia}\fourier{k}\fiak{I}{a}{k}\fiak{I}{a}{k}
 \left[k_0^2-(\kpa)^2\right]
 \mbyp\frac{|\Delta_I|^2}{4\dmi^2(\eta^2-1)}\label{Seff2}\;,
\end{equation}
%\end{widetext}
where  $\kpa=\unitqia(\vk\cdot\unitqia)$ and
$\fiak{I}{a}{k}=\vu_I\cdot(2\qia)$. Inverse Fourier
transforming back to position space, and taking out a factor of half for future
convenience, we obtain the effective action
for the displacement fields:
\begin{equation}
{\cal{S}}[{\bf u}]=\ha\intspace{x}
 \sum_I\mbyp\frac{2|\Delta_I|^2\eta^2}{(\eta^2-1)}\sum_{\qia}\left[
 \partial_0(\unitqia\cdot\vu_I)\partial_0(\unitqia\cdot\vu_I) -
 (\unitqia\cdot\vec{\partial})(\unitqia\cdot\vu_I) (\unitqia\cdot\vec{\partial})(\unitqia\cdot\vu_I)
 \right]\label{Seff3}\;.
\end{equation}
\end{widetext}
This is the low energy effective action for phonons in any crystalline color
superconducting phase, valid to second order in derivatives, to
second order in the gap parameters $\Delta_I$'s and to second order in
the phonon fields $u_I$.    This is the central technical result of our paper.

Because we are interested in long wavelength, small amplitude, phonon
excitations, expanding to second order in derivatives and to second order
in the $u_I$ is satisfactory in every respect.  Not so for the expansion
to second order in the $\Delta_I$.  As we have discussed previously,
the Ginzburg-Landau approximation is at the point of breaking down
in the most favorable CubeX and 2Cube45z crystal structures.
Before proceeding, we therefore ask
what kind of
corrections to (\ref{Seff2}) will arise at higher order in $\Delta$.
The first thing
to note is that in the weak coupling limit $\mu$ appears only as an
overall factor of $\mu^2$ in front of the fermion loop integrals.
After simplifying the fermionic propagators as in (\ref{prop1}) and
(\ref{prop2}) and taking $\Lambda$ to $\infty$, the only two independent
dimensionful quantities that remain in the integrals are $k$ and
$\dm{I}$. (Recall that $|\q{I}{}|$ is
given by $\eta\dm{I}$ and so is not independent.) Since we found the action only up
to terms which are second order in the derivatives and second order
in $\Delta$, to ensure the Lagrangian density has dimension four,
only a dimensionless factor can multiply
$\mu^2|\Delta_I|^2\partial^2 \vu_I^2$, as we can see is true in
Eq.~(\ref{Seff3}). Higher powers of $\Delta^2$
will appear in Eq.~(\ref{Seff2}) in combination
with higher compensating powers of $\dm{}^{-2}$. Consequently, there
will be corrections to the coefficients of $k_0^2$ and $(\kpa)^2$ in
(\ref{Seff2}) suppressed by factors of  $(\Delta^2/\dm{}^2)$ relative to
the leading order result that we have obtained.
In addition, there will be new terms. There is
no reason to expect that the coefficient of  $(\kpe)^2$
will remain zero at ${\cal{O}}(\mu^2|\Delta_I|^4(\partial
\vu_I)^2/\dm{}^2)$. Finally, we see that there are no terms in (\ref{Seff2}) 
that ``mix'' the different $\vu_I(k)$. This follows
from the color-flavor structure of the condensate as discussed
above. At higher order, there will be terms proportional to
$\mu^2|\Delta_I\Delta_J|^2 \partial\vu_I\partial\vu_J/\dm{}^2$,
which do ``mix'' the different $\vu_I$'s.

With the phonon effective action now in hand, in Section IV we shall
relate the coefficients
of the terms in ${\cal{S}}(\vu)$ involving spatial derivatives acting on the
displacement fields to the shear modulus of crystalline color
superconducting quark matter with specified crystal structures.

\section{Extracting the shear modulus\label{linear response}}

 We see from Eq.~(\ref{Seff3}) that the action of the phonon fields, ${\cal{S}}(\vu)$,
 is a sum of two terms: the
kinetic energy, which has time derivatives
acting on the fields $\vu$, and the potential energy,
which has
spatial derivatives acting on $\vu$. From the basic theory of
elastic media \cite{Landau:Elastic}, the potential energy is related
to the elastic moduli that describe  the energy cost of small deformations
of the crystal.
In
this Section, we present this relation explicitly and calculate the
shear modulus for the CubeX and 2Cube45z crystal structures.

\subsection{Generalities\label{linear response generalities}}

 Let there be a set of displacement fields $\vu_{I}$
 propagating in a crystalline color superconducting material.
 (We will set the problem up in the general case where
 all the $\Delta_I$ are nonzero, meaning that
$I$ runs from $1$ to $3$.)
The kinetic energy density
for the displacement fields  takes the form
\begin{equation}
{\cal{K}}\,=\,\ha\sum_{IJ}\sum_{mn}\rho^{mn}_{IJ}\,(\partial_0
\vu_I^m)(\partial_0 \vu_J^n) \label{general ke}\;,
\end{equation}
where $\vu^m_I$, $\vu^n_J$ are the space components of the vectors
$\vu_I$ and $\vu_J$ respectively. (We will use the indices $m$, $n$,
$u$ and $v$ to represent spatial indices in the following).  As
we are working only to order $\Delta^2$, the only nonzero components of
$\rho^{mn}_{IJ}$  are those with $I=J$.  We will  choose the direction
of the axes, $x$, $y$ and $z$ such that for every $I$ and $J$,
$\rho_{IJ}^{mn}$ is diagonal in the $m$ and $n$ indices and we will
denote the diagonal components of $\rho^{mn}_{IJ}$ by
$\rho^{m}_{I}$.
We can then rewrite the kinetic energy density as
\begin{equation}
{\cal{K}}=\ha\sum_I\sum_m \rho_I^m (\partial_0 \vu_I^m)(\partial_0
\vu_I^m) \label{diagonal ke}\;.
\end{equation}
At higher orders in $\Delta^2$, we could need to
choose a new linear combination of fields $\tilde{\vu}_I^m =
A_{IJ}^{mn} \vu_J^n$ to render the kinetic energy diagonal in the $IJ$
and $mn$ indices.

The potential energy  density to quadratic order in the displacement
fields can be written as
\begin{equation}
{\cal{U}}=\ha\sum_{IJ}\sum_{{mn}\atop{uv}}\lambda_{IJ}^{munv}
\frac{\partial \vu_I^m}{\partial x^u} \frac{\partial
\vu_J^n}{\partial x^v} \label{general pe}\;,
\end{equation}
where $\lambda_{IJ}^{munv}$ is the elastic modulus tensor. The
components  of the tensor ${\partial \vu_I^m}/{\partial x^n}$
that are antisymmetric in the $mn$ space indices are related to
rigid rotations. The symmetric components of the tensor, namely the
``strain tensor''
\begin{equation}
s_I^{mu}=\ha\Bigl(\frac{\partial \vu_I^m}{\partial
x^u}+\frac{\partial \vu_I^u}{\partial x^m}\Bigr)\label{strain}\;,
\end{equation}
tell us about deformations of the medium.
In the  previous Section we have
shown that,  to order $\Delta^2$, there is no interaction between
the displacement fields ${\bf u}_I$ and ${\bf u}_J$ with $I$
different from $J$. Therefore $\lambda_{IJ}^{munv}$ is diagonal in
the $I$ and $J$ indices and, denoting the diagonal entries by
$\lambda_I^{munv}$, we find
\begin{equation}
{\cal{U}}=\ha\sum_{I}\sum_{{mn}\atop{uv}}\lambda_{I}^{munv}
s_I^{mu}s_I^{nv}\label{diagonal pe}\;.
\end{equation}

Next, we define the stress tensor acting on the crystal $I$ as~\cite{Landau:Elastic}
\begin{equation}
\sigma_I^{mu} = \frac{\partial {\cal{U}}}{\partial s_I^{mu}}\;,
\label{stress1}
\end{equation}
which is symmetric in its spatial indices.
For a potential ${\cal U}$ that is quadratic in the displacement
fields and is given by (\ref{diagonal pe}), the stress tensor is
\begin{equation}
\sigma_I^{mu}
=
\lambda_I^{munv}s_I^{nv}\label{stress2}\; .
\end{equation}
The diagonal components of $\sigma$ are proportional to the
compression exerted on the system and are therefore related to  the
bulk  modulus of the crystalline color superconducting
quark matter. Since unpaired quark
matter  has a pressure $\sim \mu^4$, it gives a contribution to the
bulk modulus that  completely overwhelms the contribution from the
condensation into a crystalline phase, which is of order
$\mu^2\Delta^2$.  We shall therefore not calculate the
bulk modulus.  On the other hand, the response to shear
stress arises only because of the presence  of the crystalline condensate.

The shear modulus can be defined as follows. Imagine exerting a
static external stress $\sigma_I$ having only an off-diagonal
component, meaning  $\sigma^{mu}_I\neq 0$ for a pair of space
directions $m\neq u$, and all the other components of $\sigma$ are
zero. The system will respond with a strain $s_I^{nv}$ satisfying
(\ref{stress1}). The shear modulus in the $mu$ plane,
$\nu_I^{mu}$, is defined as half the ratio of the stress to the strain:
\begin{equation}
\nu_I^{mu} = \frac{\sigma_I^{mu}}{2s_I^{mu}}\label{define shear
modulus}\;,
\end{equation}
where the indices $m$ and $u$ are not summed.  For a quadratic potential,
with $\sigma_I^{mn}$ given by (\ref{stress2}), the shear modulus is
\begin{equation}
\nu_I^{mu} = \frac{\lambda_I^{munv} s_I^{nv}}{2s_I^{mu}}\label{quadraticshearmodulus}\;,
\end{equation}
where $n$ and $v$ are summed but $m$ and $u$ are not.
For all the crystal structures that we shall consider below,  the
only nonzero entries in $\lambda^{munv}$ with $m\neq u$ are
the
$\lambda^{mumu}$ entries, meaning that (\ref{stress2}) simplifies even
further to
\begin{equation}
\nu_I^{mu} = \ha\lambda_I^{mumu} \label{simplifiedshearmodulus}\;,
\end{equation}
again with $m$ and $u$ not summed.

Putting Eq.~(\ref{diagonal ke}) and Eq.~(\ref{diagonal pe}) together,
the action for the displacement fields can be written as
\begin{widetext}
\begin{equation}
{\cal S}[{\bf u}]=\intspace{x}({\cal{K-U}}) = \ha\intspace{x}\left(
     \sum_I\sum_m \rho_I^m (\partial_0 \vu_I^m)(\partial_0 \vu_I^m)
     -\sum_{I}\sum_{{mn}\atop{uv}}\lambda_I^{munv}
     s_I^{mu}s_I^{nv}\right)
\label{full action}\;.
\end{equation}
\end{widetext}
The equations of motion
obtained by extremizing the action ${\cal S}[{\bf u}]$ with respect to
the displacement fields ${\bf u}$ are
\begin{equation}
\rho_I^m \frac{\partial^2\vu_I^m}{\partial t^2} = \lambda_I^{munv}
\partial_u\partial_n u_I^v\label{eom}\;,
\end{equation}
where $I$ and $m$ are not summed.
The dispersion relations are found by solving
\begin{equation}
{\rm Det}\bigl[\rho_I^m k_0^2\delta_{mn} - \lambda_I^{munv} k_u k_v
\bigr] = 0
\label{DispersionRelation}
\end{equation}
for all $I$, where $m$ is again not summed.

\subsection{Elastic moduli of crystalline phases}

In order to set up the extraction of the elastic moduli of crystalline
phases, we need to rewrite the action (\ref{Seff3}) for a generic
crystalline phase in a form which makes comparison to (\ref{full action})
straightforward.
Writing the spatial indices in Eq.~(\ref{Seff3}) explicitly, we
obtain the low energy phonon effective action in the form
\begin{widetext}
\begin{equation}
%\begin{split}
S[{\bf u}]=%&
\ha\intspace{x}\sum_I \kappa_I \left[
  \left(
% \frac{2 \mu^2|\Delta_I|^2\eta^2}{\pi^2(\eta^2-1)}
    \sum_{\qia}(\hat{q}_I^a)^m(\hat{q}_I^a)^n \right)(\partial_0\vu_I^m)(\partial_0\vu_I^n)
    %\\
 %&
 -\left(
% \frac{2 \mu^2|\Delta_I|^2\eta^2}{\pi^2(\eta^2-1)}
    \sum_{\qia}(\hat{q}_I^a)^m(\hat{q}_I^a)^u(\hat{q}_I^a)^n(\hat{q}_I^a)^v\right)
    (\partial_u\vu_I^m)(\partial_v\vu_I^n)
 \right]\label{Seff4}\;
%\end{split}
\end{equation}
\end{widetext}
where we have defined
\begin{equation}
\kappa_I\equiv
\frac{2\mu^2|\Delta_I|^2\eta^2}{\pi^2(\eta^2-1)}\;.
\label{lambdapw}
\end{equation}
For a given crystal structure, upon evaluating the sums in (\ref{Seff4}) and
then using the definition (\ref{strain}) to compare (\ref{Seff4}) to
(\ref{full action}), we can extract expressions for the $\lambda$ tensor
and thence for the shear moduli.
The quantity $\kappa$ is
related to the elastic modulus for a condensate whose ``crystal'' structure
is just a single plane wave, as discussed in Appendix~\ref{single pw}.

In the next two Subsections, we will calculate  the shear modulus for the CubeX and 2Cube45z
crystals.
We will not discuss the expression for the
kinetic energy density, ${\cal{K}}$, but it is easy (and necessary) to check that in each
case below we have
chosen our axes such that ${\cal{K}}$  only contains terms that are diagonal
in the spatial indices $m$ and $n$.

Note that henceforth we set
$\Delta_1=0$ and $\Delta_2=\Delta_3=\Delta$, meaning that
\begin{equation}
\kappa_2=\kappa_3\equiv\kappa=\frac{2\mu^2|\Delta|^2\eta^2}{\pi^2(\eta^2-1)}
\simeq 0.664\,\mu^2|\Delta^2|\;.\label{kappavalue}
\end{equation}
The shear moduli that we evaluate each take the form of
a dimensionless constant times $\kappa$.

\subsection{Shear modulus for the CubeX crystal\label{CubeX}}

Orienting the axes as shown in the left panel of Fig.~1,
we have $\hatq{2}{}=\{(1/\sqrt{3})(\pm\sqrt{2},0,\pm 1)\}$ and
$\hatq{3}{}=\{(1/\sqrt{3})(0,\pm\sqrt{2},\pm 1)\}$. Calculating the
relevant sums and substituting in (\ref{Seff4}), we find that the
potential energy is given by
\begin{widetext}
\begin{equation}
\begin{split}
{\cal{U}} =
&\frac{4}{9}\kappa\Bigl(4(s_2^{xx})^2+(s_2^{zz})^2\Bigr)
+\frac{16}{9}\kappa(s_2^{xx}s_2^{zz})
+\frac{16}{9}\kappa\Bigl((s_2^{xz})^2+(s_2^{zx})^2\Bigr)\\
&+ \frac{4}{9}\kappa\Bigl(4(s_3^{yy})^2+(s_3^{zz})^2\Bigr)
+\frac{16}{9}\kappa(s_3^{yy}s_3^{zz})
+\frac{16}{9}\kappa\Bigl((s_3^{yz})^2 +(s_3^{zy})^2\Bigr)
\label{SeffCubeX}\;.
\end{split}
\end{equation}
\end{widetext}
Recall that the only components of the
stress tensor that are relevant to the calculation of the shear
modulus
are given by $\partial {\cal U}/\partial s_I^{mu}$ for $m\neq u$.
These are
\begin{equation}
\sigma_2^{xz}=\sigma_2^{zx}
%=\frac{\partial{\cal{U}}}{\partial s_3^{zy}}
=\frac{32}{9}\kappa\, s_2^{zx}\label{stress2 CubeX}
\end{equation}
and
\begin{equation}
\sigma_3^{yz}=\sigma_3^{zy}
%=\frac{\partial{\cal{U}}}{\partial s_3^{zy}}
=\frac{32}{9}\kappa\, s_3^{zy}\label{stress3 CubeX}\;,
\end{equation}
from which we obtain
\begin{equation}
\nu_2^{xz}=\nu_2^{zx}=\frac{\sigma_2^{zx}}{2s_2^{zx}}=\frac{16}{9}\kappa
\end{equation}
and
\begin{equation}
\nu_3^{yz}=\nu_3^{zy}=\frac{\sigma_3^{zy}}{2s_3^{zy}}=\frac{16}{9}\kappa\;.
\end{equation}
We can display
the result succinctly by writing two shear matrices  $\nu_2$ and
$\nu_3$, which have only off-diagonal entries and are symmetric in
the spatial indices:
\begin{equation}
\nu_2=\frac{16}{9}\kappa\left( \begin{array}{ccc}
0 & 0 & 1\\
0 & 0 & 0\\
1 & 0 & 0
\end{array}
\right)\,,\hspace{.3cm}  \nu_3=\frac{16}{9}\kappa\left(
\begin{array}{ccc}
0 & 0 & 0\\
0 & 0 & 1\\
0 & 1 & 0
\end{array}
\right)\label{nu2 and nu3}\;.
\end{equation}
The zeroes in these matrices are easily understood.
The $\Delta_2$ crystal is translation invariant in the $y$-direction,
because all the wave vectors in the set $\setq{2}{}$ lie in the $xz$-plane.
This means
that the $xy$- and $yz$-components of $\nu_2$ are zero.  The only
nonzero shear modulus is that for shear in the $xz$-plane.
Note also that the $\Delta_2$ crystal has nonzero $\lambda_3^{xxxx}$ and $\lambda_3^{zzzz}$,
meaning that it has a nonzero Young's modulus for compression or stretching in the
$x$- and $z$-directions confirming that, as the shear modulus indicates, it is rigid
against deformations in the $xz$-plane.
Similarly, the $\Delta_3$ crystal is translation invariant in the $x$-direction,
meaning that the only nonzero component of the shear
modulus $\nu_3$ is that for shear in the $yz$-plane.

The vortices in rotating crystalline color superconducting quark matter
have currents of $u$, $d$ and $s$ quark-number flowing around them,
meaning that the phase of both the $\Delta_2$ and $\Delta_3$ condensates
winds once by $2\pi$ around a rotational vortex, and meaning that both
$\Delta_2$ and $\Delta_3$ vanish at the core of the vortex. This in turn
means that it will be free energetically favorable for the vortices to
be pinned at places where the $\Delta_2$ and $\Delta_3$ condensates
already vanish in the absence of a vortex.
The $\Delta_2$ crystal has two families of nodal planes
where $\Delta_2({\bf r})$ vanishes. One class of nodal planes are
parallel to the $xy$-plane and are located at
$z=((2n+1)\pi\sqrt{3})/(4q)$, where $n$ is an integer. The others are
parallel to the $yz$-plane and are located at
$x=((2n+1)\pi\sqrt{6})/(4q)$.
Similarly, the $\Delta_3$ crystal has nodes along
$z=((2n+1)\pi\sqrt{3})/(4q)$ and $y=((2n+1)\pi\sqrt{6})/(4q)$.
So, we expect that the most
favorable location of the vortices will be within
the common nodal planes of
the $\Delta_2$ and $\Delta_3$ condensates, namely,
$z=((2n+1)\pi\sqrt{3})/(4q)$. If these vortices are oriented in the $x$-direction,
they will preferentially be located (i.e. will be pinned at)
at $x=((2n+1)\pi\sqrt{6})/(4q)$. And, if the vortices in an array
of vortices oriented in the $x$-direction try to move apart (i.e.
move in the $yz$-plane) as the rotation
slows, in order to move they will have to shear the $\Delta_2$ crystal
which has a nonzero $\nu_3^{yz}$.  Similarly, if the vortices
are oriented in the $y$-direction, they
will preferentially be located (i.e. will be pinned at)
at $y=((2n+1)\pi\sqrt{6})/(4q)$. And, if the vortices in an array
of vortices oriented in the $y$-direction try to move apart (i.e.
move in the $xz$-plane), they will have to shear the $\Delta_3$ crystal
which has a nonzero $\nu_3^{xz}$.
Thus, the nonzero shear moduli that we have found in (\ref{nu2 and nu3}) are
sufficient to ensure that vortices pinned within the CubeX phase are pinned
to a rigid structure, with the relevant shear modulus having a magnitude
$16\kappa/9$.

We note as an aside that further evidence for the rigidity of the CubeX crystal
can be found by evaluating the phonon velocities and showing that at long wavelengths
the velocity of transverse phonons
(which are found in a rigid solid but not in a fluid)
is comparable to that of the longitudinal phonons which are
found in both fluids and solids.   We will evaluate the velocities
of the longitudinal phonons upon ignoring the existence of longitudinal
oscillations in the gapless fermions, which have velocity $1/\sqrt{3}$
in the limit of weak coupling.  For this reason,
the longitudinal phonon velocity that 
we calculate should be seen only as a benchmark against which to
compare the transverse phonon velocity.
The true 
sound modes would be linear
combinations of the longitudinal phonons and the fermionic sound waves,
which must in reality be coupled.  This complication does not arise for 
transverse phonons: the fluid of gapless fermions has no transverse sound
waves; they can only arise as excitations of a rigid structure, like the
crystalline condensate we analyze.  
Consider as an example the phonons of
the $\Delta_2$-crystal.
From the dispersion relations (\ref{DispersionRelation}) it is easy to
show using
$\rho_2^x=8\kappa/3$, $\rho_2^y=0$ and $\rho_2^z=4\kappa/3$ that longitudinal phonons
propagating in the $x$-direction have $v=\sqrt{2/3}$ while transverse
phonons propagating in this
direction have  the same $v=\sqrt{2/3}$.
For propagation in the $z$-direction,
both modes turn out to have $v=\sqrt{1/3}$.
For propagation in other directions, there are two phonon modes
with differing velocities.

\subsection{Shear modulus for the 2Cube45z crystal\label{2Cube45z}}

Let us orient the coordinate axes such that
$\hatq{2}{}$ contains the eight
wave vectors ${(1/\sqrt{3})(\pm 1,\pm 1,\pm 1)}$ and
$\hatq{3}{}$ contains ${(1/\sqrt{3})(\pm \sqrt{2},0,\pm
1)}\cup{(1/\sqrt{3})(0,\pm \sqrt{2},\pm 1)}$. These wave vectors are
shown in the right panel of Fig.~1.
The potential energy is given by
\begin{widetext}
\begin{equation}
\begin{split}
{\cal{U}} =
&\frac{8}{9}\kappa\Bigl((s_2^{xx})^2+(s_2^{yy})^2+(s_2^{zz})^2\Bigr)
+\frac{16}{9}\kappa\Bigl((s_2^{xx})(s_2^{zz})+(s_2^{yy})(s_2^{zz})
            +(s_2^{xx})(s_2^{yy})\Bigr)\\
&\;\;+\frac{16}{9}\kappa\Bigl((s_2^{xy})^2+(s_2^{yx})^2+(s_2^{yz})^2+(s_2^{zy})^2+(s_2^{xz})^2+(s_2^{zx})^2\Bigr)\\
& +\frac{8}{9}\kappa\Bigl( 2(s_3^{xx})^2+2(s_3^{yy})^2+(s_3^{zz})^2\Bigr)
+\frac{16}{9}\kappa\Bigl((s_3^{yy})(s_3^{zz})+(s_3^{xx})(s_3^{zz})\Bigr)\\
&\;\;+\frac{16}{9}\kappa\Bigl((s_3^{xz})^2+(s_3^{zx})^2+(s_3^{yz})^2+(s_3^{zy})^2\Bigr)
\label{Seff2Cube45z}\;,
\end{split}
\end{equation}
\end{widetext}
from which one can read off the nonzero entries of the $\lambda_I^{munv}$
tensors for $I=2$ and $I=3$.
In the case of $\Delta_2$, where the axes are oriented perpendicular to
the nodal planes of the crystal,
the form of $\lambda_2^{munv}$ for the cubic crystal are easily inferred
from the symmetries of the cube~\cite{Landau:Elastic}.  There are
in general only three independent
nonzero entries in $\lambda_2^{munv}$, corresponding to the terms
read from (\ref{Seff2Cube45z})
with the form $\lambda_2^{mmmm}$,
$\lambda_2^{mmnn}$ and
$\lambda_2^{mumu}$. The form of $\lambda_2^{munv}$ read from (\ref{Seff2Cube45z})
is therefore valid to all orders in $\Delta_2$, although of course the values
of the coefficients, including in particular the equality between the
$\lambda_2^{mmnn}$ and
$\lambda_2^{mumu}$ coefficients, will receive corrections at higher order.
Finally, note that $\lambda_3^{munv}$ read from (\ref{Seff2Cube45z}) is
obtained from $\lambda_2^{munv}$ by rotating this tensor by 45$^\circ$ about the $z$-axis.
Note that $\lambda_3^{xyxy}=\lambda_3^{yxyx}$ vanishes. This is a consequence
(after the 45$^\circ$ rotation) of the equality of
$\lambda_2^{mmnn}$ and $\lambda_2^{mumu}$ in the $\Delta_2$ crystal, and
is therefore not expected to persist at higher order in $\Delta_3$.

As in the previous subsection, we extract the $\sigma_I$
tensors and the matrices of shear moduli $\nu_I$ from the potential ${\cal U}$
of (\ref{Seff2Cube45z}),
obtaining in this case
\begin{equation}
\nu_{2}=\frac{16}{9}\kappa\left( \begin{array}{ccc}
0 & 1 & 1\\
1 & 0 & 1\\
1 & 1 & 0
\end{array}
\right)\,,\ \
\nu_{3}=\frac{16}{9}\kappa\left( \begin{array}{ccc}
0 & 0 & 1\\
0 & 0 & 1\\
1 & 1 & 0
\end{array}
\right)\label{nu 2Cube45z}\;.
\end{equation}
As discussed above, there is no symmetry reason for $\nu_3^{xy}=0$, so we expect that
this component of the shear modulus is nonzero at order $\Delta^4/\delta\mu^2$.
Note that if we were to rotate our coordinate axes by 45$^\circ$ about the
$z$-axis, it would be $\nu_3$ that has the zero entry while $\nu_2$ would have
all off-diagonal entries nonzero. (To confirm this, rotate the $\lambda_I^{munv}$
tensors and re-extract the $\nu_I$ matrices, which are not tensors.)

The $\Delta$ crystal has nodes along $x=((2n+1)\pi\sqrt{3})/(4q)$,
$y=((2n+1)\pi\sqrt{3})/(4q)$ and $z=((2n+1)\pi\sqrt{3})/(4q)$. The $\Delta_3$
crystal has nodes along $x\pm y =((2n+1)\pi\sqrt{6})/(4q)$ and $z=((2n+1)\pi\sqrt{3})/(4q)$.
The nodes common to
both lie along the $z=((2n+1)\pi\sqrt{3})/(4q)$ planes. We therefore expect that
the crystal will orient itself relative to the rotation axis such that rotation vortices
lie within these planes.  Depending on their orientation within the planes, they
could be pinned where the perpendicular nodal planes of either the $\Delta_2$
or the $\Delta_3$ crystals intersect the $z=((2n+1)\pi\sqrt{3})/(4q)$ planes.

We learn from our analysis that the crystals are weaker (smaller shear modulus)
with respect to shear in certain planes. We saw this explicitly for $\nu_3^{xy}$, which
is zero to order $\Delta_3^2$ and thus presumably weaker although nonzero when
higher order terms are included.  The same will apply to shear
in any plane obtained from this one by a symmetry transformation of the
crystal, and will apply to the analogous planes for the $\Delta_2$ crystal. Note, however,
that in the 2Cube45z structure  the weak planes for
the $\Delta_2$ and $\Delta_3$ crystals do not coincide.  This means
that if it so happens that motion of a rotational vortex in a certain
direction is only impeded
by the weaker shear modulus of the $\Delta_2$ crystal, it will in fact
be obstructed by the stronger shear modulus of the $\Delta_3$ crystal, or vice versa.
Thus, the relevant shear modulus in the analysis of vortex pinning and
pulsar glitches is the stronger one, which we find to be $16\kappa/9$ to
order $\Delta^2$.

As in Section IV.C, we can find further evidence for the rigidity of the 2Cube45z
crystal by evaluating the velocity of the transverse
and longitudinal phonons.
Considering the $\Delta_2$-crystal as an example,
from the dispersion relations (\ref{DispersionRelation}) and
$\rho_2^x=\rho_2^y=\rho_2^z=8\kappa/3$ we find that
for propagation in the $x$- or $y$- or $z$-direction the
longitudinal phonon mode 
and the two transverse phonon modes all have $v=\sqrt{1/3}$.
For propagation in the $x\pm z$ directions, the longitudinal mode
has $v=\sqrt{2/3}$ while one transverse mode has $v=\sqrt{1/3}$
and the other transverse mode, corresponding to transverse oscillations for which
the restoring force would be given by the component of the
shear modulus which vanishes at order $\Delta^2$, turns
out indeed to have $v=0$.  We see that the velocity of
both the longitudinal and transverse
phonons is anisotropic, as expected in a crystal, and see that
they are comparable in magnitude, confirming that 
the nonzero components of the shear moduli are as large as the
longitudinal elastic moduli, as expected for a very rigid body.

\section{Conclusion \label{conclusion}}

\subsection{The rigidity of crystalline color superconducting quark matter}

We have calculated the shear moduli of crystalline color superconducting
quark matter with the CubeX and 2Cube45z crystal structures.  Within the
Ginzburg-Landau analysis of Ref.~\cite{Rajagopal:2006ig}, one or other of these crystal
structures is favored over
unpaired quark matter and over spatially uniform paired phases like the CFL phase
in the wide regime of densities given in Eq.~(\ref{crystallineregime}).
As we have explained in Sections IV.C and IV.D, in both
these structures the components
of the shear moduli that make the crystals
rigid with respect to vortices pinned within them take on the
same value to order $\Delta^2$, given by
\begin{equation}
\nu_{\rm CQM} = \frac{16}{9}\kappa
\end{equation}
with $\kappa$ defined by (\ref{kappavalue}).
Evaluating $\kappa$ yields
\begin{equation}
\nu_{\rm CQM} = 2.47\, {\rm MeV}/{\rm fm}^3
\left(\frac{\Delta}{10~{\rm MeV}}\right)^2 \left(\frac{\mu}{400~\rm{MeV}}\right)^2
\label{nunumerical}
\end{equation}
for the shear moduli of crystalline quark matter with these two crystal structures.
If quark matter is found within neutron stars, it is reasonable to estimate that
its quark number chemical potential will lie in the range
\begin{equation}
350~{\rm MeV}<\mu<500~{\rm MeV}\ .
\label{murange}
\end{equation}
The gap parameter $\Delta$ is
less well known.
According to the Ginzburg-Landau calculations of Ref.~\cite{Rajagopal:2006ig},
$\Delta/\Delta_0$ is about $1/4$ to $1/2$, with  $\Delta_0$ the CFL gap parameter
for $M_s=0$.
Here,  $\Delta/\Delta_0$ for the CubeX crystal structure
somewhat larger than that for the 2Cube45z structure and $\Delta/\Delta_0$ is
a slowly increasing function of $M_s^2/\mu$, meaning a slowly decreasing
function of density~\cite{Rajagopal:2006ig}.
It is reasonable to estimate that
$\Delta_0$ is  between 10 and
100 MeV, but if $\Delta_0$ is in the upper half of this range then quark matter
at accessible densities is likely in the CFL phase, rather than in the crystalline
phase.  So, we suggest that in interpreting (\ref{nunumerical}) it is reasonable
to estimate that
\begin{equation}
5~{\rm MeV}\lesssim \Delta \lesssim 25~{\rm MeV}\ ,
\label{Deltarange}
\end{equation}
keeping in mind that a part of the uncertainty encompassed by this
range comes from our lack of knowledge of $\Delta_0$ and a part comes
from the $M_s^2/\mu$-dependence of $\Delta/\Delta_0$ described in
Ref.~\cite{Rajagopal:2006ig}.
The estimates (\ref{Deltarange}) and
(\ref{murange}) mean that our result
(\ref{nunumerical}) implies
\begin{equation}
0.47~{\rm MeV}/{\rm fm}^3 < \nu_{\rm CQM} < 24~{\rm MeV}/{\rm fm}^3\ .
\label{nuCQMrange}
\end{equation}
We shall take this as an estimate of the magnitude of $\nu_{\rm CQM}$,
although (\ref{nunumerical}) is a better representation of our result
for use in future work.

One qualitative way to appreciate how rigid the crystalline phases of quark
matter are
is to calculate the (anisotropic) velocities of long wavelength
transverse and longitudinal phonons, as we have done  for a
few directions of propagation in the CubeX and 2Cube45z crystal
structures in Sections IV C and IV D respectively. We find that the transverse
modes, whose restoring forces are governed by the shear moduli,
propagate with velocities that are comparable to 
the velocity of longitudinal phonons. 

To appreciate more
quantitatively how rigid the crystalline phases of quark matter prove
to be, we compare the shear modulus that we have calculated to
that for the standard neutron star crust, which is a conventional
crystal of positively charged ions immersed in a fluid of
electrons (and, at sufficient depth, a fluid of neutrons). The shear
modulus of this solid can be expressed as~\cite{Strohmayer}
\be
\nu_{\rm NM} = c\frac{n_i (Ze)^2}{a}\,,
\ee
where $n_i$ is the number density of ions in the crust, $Z$ is the
atomic number of the positively charged ions,
$a= (3/(4 \pi n_i))^{1/3}$ is the
average inter-ion spacing, $e^2 \simeq 4\pi/137$ and
$c\sim 0.1 - 0.2$ is a dimensionless constant.
Because the crust is electrically neutral, the number density
of ions is related to $n_e$, the electron number density, by
$n_i = n_e/Z$.  And, $n_e$ is given in terms of the mass and
electric chemical potential $\mu_e$ of the electrons by
\begin{equation}
n_e = \frac{(\mu_e^2-m_e^2)^{3/2}}{3 \pi^2}\,,
\end{equation}
where $\mu_e$ is
estimated to be in the range $20-80$ MeV and $Z\sim 40-50$~\cite{Strohmayer}.
Using these estimates, we find
\begin{equation}
0.092~{\rm keV}/{\rm fm}^3 < \nu_{\rm NM}<23~{\rm keV}/{\rm fm}^3 \ .
\label{nuNMrange}
\end{equation}
Comparing to (\ref{nuCQMrange}), we see that crystalline quark matter
is more rigid than the conventional neutron star crust by at least a factor of 20,
and possibly by about three orders of magnitude.

We conclude that crystalline color superconducting quark matter
is a very good solid indeed, which is remarkable since it is
at the same time superfluid.

\subsection{Toward pulsar glitch phenomenology}

As discussed in the Introduction, the glitches that are observed to
interrupt the gradual spin-down of spinning neutron stars are thought
to arise from the sudden unpinning of an array of rotational vortices that had been
pinned in place, at a fixed area density and hence
a fixed angular momentum, while the other components of the star
and in particular the observed surface had been gradually slowing down.
When the stressed vortices unpin and separate, the superfluid component
loses angular momentum while the surface spins up.    Can these phenomena
originate within crystalline color superconducting quark matter in the core
of a neutron star?   This phase of matter is a superfluid while at the
same time having a rigid spatial modulation of its superfluid condensate,
as we have seen.   Understanding whether this makes it a plausible
locus for the origin of pulsar glitches requires addressing three
questions: Is crystalline color superconducting quark matter rigid enough?
Do vortices in this phase of matter get pinned? And, how rapidly can angular
momentum be transferred from a crystalline quark
matter core that has just glitched to the outer crust
whose surface is observed?

Our calculation constitutes an
affirmative answer to the first question.  We have shown
that both the CubeX and 2Cube45z crystal structures have
shear moduli with magnitude (\ref{nunumerical})
which are 20 to 1000 times greater than those of the
conventional neutron star crust within which glitches
have long been assumed to originate.

Next, do vortices in fact get pinned? With what pinning
force?  This is a much harder question
to address quantitatively because doing so requires going beyond
the long wavelength phonon effective action.  The question is
what is the difference in the energy per unit length of
a  vortex centered on a nodal plane (or at the intersection
of two nodal planes) of the condensate and one centered
half way between neighboring nodal planes.
Understanding this quantitatively requires constructing a vortex
solution in the crystalline background, which is
a challenging task.  In the conventional neutron star crust,
a vortex in a neutron superfluid is pinned on ``impurities'' embedded
in the superfluid, namely the lattice of positively charged nuclei.
In rotating crystalline color superconducting quark matter, the vortices
are deformations of the phase and  magnitude of the
same condensate whose underlying magnitude modulation is
the origin of the pinning.
Unlike in the case of the shear modulus, which describes the response
to a stress on length scales long compared
to those characteristic of the crystal itself, the deformations introduced
by a vortex will occur on length scales comparable to the lattice spacing
of the underlying crystal.
This means that constructing
the vortices must be done self-consistently with analyzing the crystal
structure itself --- the pinning sites are in no sense extraneous impurities.
We can provide a crude estimate of the pinning force, but
we defer a quantitative response to this challenge to future work.

To estimate the pinning force, let us suppose (contrafactually) that
the core radius of a vortex
$\xi\sim1/\Delta$ is much smaller
than the spacing between nodal planes of the crystalline condensate.
If such a vortex is located where the underlying crystalline condensate
is maximal, it will have to deform that condensate maximally since
at the center of the vortex the condensate must vanish. Clearly, it
will be energetically advantageous to locate the vortex at the
intersection of nodal planes where the condensate already
vanishes in the absence of a vortex. This argument translates into
a pinning energy per unit length given at the level of dimensional analysis by
\begin{equation}
\frac{E_p}{\ell} = f\, |\Omega_{\rm crystalline}|\, \xi^2
\end{equation}
where $|\Omega_{\rm crystalline}|$ is the condensation energy
of the crystalline phase and
where $f$ is some dimensionless factor. The corresponding pinning
force per unit length is given by
\begin{equation}
\frac{F_p}{\ell} = \frac{f\, |\Omega_{\rm crystalline}|\, \xi^2}{b}\ ,
\end{equation}
where the length scale
$b$ is half the spacing between  neighboring nodal planes
and hence one quarter of the lattice spacing.
In both the CubeX and 2Cube45z crystals,
$b=\pi\sqrt{3}/(4q)=1.13 /\delta\mu$.    Recalling that $\delta\mu=M_s^2/(8\mu)$,
we can get a sense of the scale of $b$ by
seeing that $b=18,12$~fm for $M_s^2/\mu=100,150$~MeV.
Reading from plots in Ref.~\cite{Rajagopal:2006ig},
we see that for $\Delta_0=25$~MeV this range of $M_s^2/\mu$ corresponds to
a robust crystalline phase with $|\Omega_{\rm crystalline}|\sim 2 \times 10^5$~MeV$^4$
and $\Delta\sim 5-10$~MeV if the crystal has the 2Cube45z structure or
$\Delta\sim 10-15$~MeV if the crystal has the CubeX structure.  We immediately
see that $\xi=1/\Delta$ and $b$ are comparable length scales, which
makes this analysis unreliable at a quantitative level.  One way of saying
this is that the dimensionless factor $f$ must then be very much less than one,
since the energy benefit by moving the vortex by a distance $b$ is
of order $|\Omega_{\rm crystalline}|\xi^2$ only  if making this move shifts
the core from a place where the condensate was maximal within the
core area $\xi^2$  to a place where it is close to vanishing.
A calculation of $f$ in the case where $\xi\sim b$ as is relevant
in our context requires a quantitative analysis, but it is clear
that the energy benefit of moving the vortex by a distance $b$
must then be $\ll |\Omega_{\rm crystalline}|\xi^2$.
Putting the pieces together, we can write  an estimate
of the pinning force per unit length as
\begin{widetext}
\begin{equation}
\frac{F_p}{\ell}=0.7 \,\frac{{\rm MeV}}{(10 {\rm fm})^2}
\left(\frac{f}{0.01}\right)
\left(\frac{|\Omega_{\rm crystalline}|}{2\times 10^5~{\rm MeV}^4}\right)
\left(\frac{\xi}{20~{\rm fm}}\right)^2
\left(\frac{15~{\rm fm}}{b}\right)\ ,
\label{pinningforceestimate}
\end{equation}
\end{widetext}
where our choice of $f\sim 0.01$ as a fiducial value is a pure guess
and the dependence of the other quantities
in the estimate (\ref{pinningforceestimate}) on $\Delta_0$
and $M_s^2/\mu$ can be obtained from 
the results of Ref.~\cite{Rajagopal:2006ig},
with the fiducial values we have used being reasonable for
$\Delta_0=25$~MeV and $M_s^2/\mu=100-150$~MeV.

We can compare our estimate (\ref{pinningforceestimate})   to the
pinning force on neutron vortices in a conventional
crust~\cite{Negele:1971vb}, in which neutron superfluid vortices are
pinned on 
nuclei~\cite{Anderson:1975,Alpar:1977,Alpar:1984a,Alpar:1984b,AlparLangerSauls:1984,Alpar:1985,Epstein:1992,Link:1993,Jones:1993,Alpar:1993,Alpar:1994,Alpar:1996,Pines:1985}.
The pinning energy of a vortex per ion on which it is pinned 
is $E_p^{\rm NM}\sim 1-3$~MeV~\cite{Alpar:1977,Alpar:1984a,Alpar:1984b}, the ions are spaced
by a lattice spacing 
$b_{\rm NM}\sim 25-50$~fm~\cite{Alpar:1984b}, and the superfluid vortices
have core radii $\xi_{\rm NM}\sim 4-20$~fm~\cite{Alpar:1984b}. Hence,
the pinning force per unit length is~\cite{Alpar:1984a,Alpar:1984b}
\begin{equation}
\frac{E_p^{\rm NM}}{b_{\rm NM}\xi_{\rm NM}}=\frac{1-3~{\rm MeV}}{(25-50~{\rm fm})
(4-20~{\rm fm})}.
\end{equation}
Although our estimate (\ref{pinningforceestimate}) is quite uncertain, given
that we have not constructed the vortex solutions for rotating  crystalline
color superconducting quark matter,  it seems reasonable to estimate that
the pinning force per unit length on vortices within a
putative crystalline quark matter
neutron star core is comparable to that on neutron superfluid vortices
within a conventional neutron star crust.

Recent calculations of the profile of vortices in BCS-paired
superfluid gases of ultracold fermionic atoms~\cite{Randeria} may
make it easier to estimate the pinning force on vortices in
the crystalline quark matter phase.  In the cold atom context,
it turns out that the radius of the vortex core is much smaller
than the correlation length $\xi\sim 1/\Delta$ which controls
the long distance form of the vortex profile. Instead, the vortex
core radius is $\sim 1/k_F$, controlled by the Fermi momentum rather
than by the correlation length.  
If this result were to be obtained
in our context, it could mean replacing $\xi$ 
in (\ref{pinningforceestimate}) by $(1/\mu)\sim 0.5$~fm, reducing
our fiducial estimate (\ref{pinningforceestimate}) by a factor
of 1600.  However, if the vortex cores do turn out to be as 
narrow as this then the assumption with which we began our 
estimate, namely
that the core size is much less
than the lattice spacing,
becomes factual rather than contrafactual.  This would
mean that there is no longer any reason to expect the dimensionless
factor $f$ to be much smaller than 1, and would
considerably reduce the uncertainty in the estimate.  
Replacing $f$ by 1 would increase
the estimate (\ref{pinningforceestimate}) by a factor of 100, resulting
in a pinning force which is only slightly smaller that on the
superfluid neutron vortices within a conventional neutron star crust.
We leave the determination of the profile of vortices in 
crystalline quark matter to future work, but it will clearly be
very interesting to see whether they have narrow
cores as in Ref.~\cite{Randeria}, and if so whether their pinning
turns out to be 
controlled by their core radii or by the correlation length.

The third question which must be addressed is how, and how quickly,
angular momentum can be transferred from a crystalline quark matter
core to the outer crust.   Some glitches are known to occur on
timescales of minutes which means that if a glitch occurs within the
core angular momentum must be transferred to the observed crust at
least this fast. The core and crust are linked via being bathed in
the same electron fluid and via magnetic fields. In the conventional
glitch scenario, when the neutron superfluid in the crust suddenly
slows down as its vortices come unpinned and the nonsuperfluid
component of the crust, which includes the ions and the electrons,
speeds up, the electron fluid couples the crust to the core well
enough that the core also speeds up within
seconds~\cite{Easson:1979,AlparLangerSauls:1984}. We therefore expect that if a glitch
occurs within a crystalline quark matter core, with this superfluid
component slowing down, 
and if moving vortices can
impart angular momentum to the electrons, then 
the electron fluid will ensure that the
entire rest of the star including the outer crust speeds up.
In the conventional
scenario, the mechanism by which moving vortices exert a torque on
the ions, and hence the electrons, in the crust has been
described in Refs.~\cite{Epstein:1992,Link:1993,Jones:1993}. 
In our case, we have not demonstrated how
moving vortices in the crystalline phase can torque up the fluid of
gapless charged quark quasiparticles, and hence the electrons.

So, tallying the status of the three questions that must be addressed:
The first is settled, answered in the affirmative by our calculation
of the shear modulus of the crystalline phase. The
second remains to be addressed quantitatively but seems to be
answered in the affirmative by our crude estimate
of the pinning force on vortices in the crystalline phase.  The third
remains open, not yet addressed in a satisfactory fashion but nevertheless
with no reason to doubt that its answer is also affirmative.  Addressing
the third question and addressing the second question quantitatively
both require constructing the rotational vortex solutions for rotating
crystalline quark matter. This is therefore the crucial remaining
step in completing the connection between the microphysics
of crystalline color superconducting quark matter and the phenomenology
of pulsar glitches, and hence determining whether the characteristics
of observed glitches rule out, or are consistent with, the presence of
crystalline color superconducting quark matter within neutron stars.

It is also worth asking whether a ``core-quake'' scenario
could be a viable model of glitches, qualitatively distinct
from that based upon vortex pinning~\cite{Pines:1972,LinkPrivate,Xu:2003xe}.
As a spinning neutron star slows down, it becomes less oblate.  This
will require macroscopic adjustments to the shape of a putative
crystalline quark matter core.  Given the enormous shear moduli of
this rigid phase of matter, enormous amounts of elastic energy would
be stored as the core is deformed and stressed, energy which would
be released in core-quakes during which the crystalline  core
``breaks'' and rearranges its structure so as to reduce its moment
of inertia, consequently increasing its angular velocity.  The
original ``crust-quake'' model for pulsar
glitches~\cite{Ruderman:1969} failed because it failed to describe
the magnitude and frequency of glitches in the Vela
pulsar~\cite{Alpar:1984b,Alpar:1993,Alpar:1994,Pines:1985}. 
Now that we know that crystalline quark
matter has shear moduli which are 20 to 1000 times larger than those
of the crust, core-quakes are worth re-investigating.   

Finally, the advent of gravity wave detectors opens a new possibility for
unique signatures of the presence of rigid matter within neutron stars,
independent of transient phenomena like 
glitches~\cite{Owen:2005fn}. LIGO has already set limits
on the steady-state gravity wave emission from 78 nearby
pulsars~\cite{Abbott:2007ce}, and these limits will become more stringent.
Owen's work~\cite{Owen:2005fn} shows that if an entire neutron star were solid,
with a shear modulus as large as that we find for 
crystalline color superconducting quark matter, it could
in principle support a quadrupole moment large enough such
that the resulting gravity waves would already have been 
detected.  This suggests that as the observational
upper limits improve, the size of a putative rigid crystalline color
superconducting quark matter core could be constrained.  
However,
Owen's estimates for a star that is rigid in its entirety cannot be
applied straightforwardly 
to the case where there is a rigid core surrounded
by a fluid ``mantle''. Oblateness about the rotational $z$-axis
is not enough to generate gravity waves; the quadrupole moment must
be mis-aligned, such that the moment of inertia
tensor satisfies $I_{xx}\neq I_{yy}$.  It is hard to imagine
how this could come about for a rigid quark matter core, whose shape
will equilibrate to follow an equipotential surface via converting
core material into mantle material or vice versa as needed at different
locations along the core/mantle interface.
Nevertheless, this line of enquiry warrants careful investigation.

\begin{acknowledgments}
We thank M.~Alford, J.~Bowers, E.~Gubankova,
B.~Link, C.~Manuel, B.~Owen, D.~Son and R.~Xu for useful discussions.
The work of MM has been supported
by the ``Bruno Rossi" fellowship program. The authors acknowledge
the hospitality of the Nuclear Theory Group at LBNL.  This research
was supported in part by the Office of Nuclear Physics of the Office
of Science of the U.S.~Department of Energy under contract
\#DE-AC02-05CH11231 and cooperative research agreement
\#DF-FC02-94ER40818.
\end{acknowledgments}

\appendix
\section{Phonon mass is zero to all orders in $\Delta$\label{mass zero}}

In this Appendix, we show explicitly that
%to one loop order
there cannot be any term in the phonon effective action
at any order in $\Delta$ and at any order in $\vu_I(x)$ which is nonzero
if $\vu_I(x)$ is  constant over space and time, other than a trivial
constant.  Equivalently, we show that if
we expand the Lagrangian density as a power series in $\vu_I(x)$,
every term has at least one derivative acting on each $\vu_I(x)$, meaning
that the mass of the phonons is zero.  This is guaranteed by
Goldstone's theorem, but the explicit demonstration of this ``obvious'' result
is nontrivial and for this reason we give it here.

Before proceeding, we recall
 Eqs.~(\ref{cond phon0}) and (\ref{cond phon1}),
which imply that
\begin{equation}
\Delta_I^u(x)=\Delta_I\sum_{\q{I}{a}}\exp\bigl(2i\q{I}{a}\cdot(\rr-\vu_I(x)\bigr)\ .
\end{equation}
Since we want to prove the result to arbitrary powers in the $\vu$
fields, we do not make an expansion in small $\vu$. In terms of
Feynman diagrams, this means that in evaluating the effective action
in Eq.~(\ref{Z2}) we resum the vertices with increasing powers of
$\vu_I$. Therefore any vertex for $\Delta_I$ comes with a factor
$\coleps\flaeps\exp\bigl(-2i\q{I}{a}\cdot\vu_I(x)\bigr)$
and a
momentum insertion $2\q{I}{a}$. On the other hand,  the vertex for
$\Delta_I^*$ comes with a vertex factor
$\coleps\flaeps\exp(2i\q{I}{a}\cdot\vu_I(x))$ and a momentum
insertion $-2\q{I}{a}$.

Integrating out the $\chi$ fields in Eq.~(\ref{Z2}) is equivalent to
calculating all possible one-fermion-loop diagrams with
arbitrarily many external phonon insertions.
(Higher loop diagrams with internal phonon propagators
are suppressed by powers of $\Lambda_{\rm IR}/\Delta$, with
$\Lambda_{\rm IR}$ the typical energy of the phonon fields,
and are therefore completely negligible.)
A generic one-loop diagram will have $n_I$
vertices proportional to
$\Delta_I \exp\bigl(-2i\q{I}{a_\kappa}\cdot\vu_I(x)\bigr)$ at which
momenta
$2\q{I}{a_\kappa}$ are inserted into the loop, and $n_I$
vertices proportional to
$\Delta_I^* \exp\bigl(2i\q{I}{a_\tau}\cdot\vu_I(x)\bigr)$ at which
momenta
$-2\q{I}{a_\tau}$ are inserted into the loop.
Here, $\kappa=1,2,3...n_I$ and $\tau=1,2,3..n_I$.
The number of appearances of $\Delta_I$ has to be
the same as that of $\Delta_I^*$ because the diagram can only depend
on powers of $|\Delta_I|^2$.  Different one-loop diagrams  correspond
to different choices of $n_1$, $n_2$ and $n_3$ and different choices
of the $\q{I}{a_\kappa}$'s and the $\q{I}{a_\tau}$'s from among the
sets $\setq{I}{}$.
The color and flavor indices for the propagators in the diagram
linking the vertices into a loop
are chosen to
be consistent with the color and flavor epsilon symbols associated to each
vertex. The contribution from a generic one-loop diagram  will be
\begin{widetext}
\begin{equation}
{\cal I} \propto
|\Delta_1|^{2n_1}|\Delta_2|^{2n_2}|\Delta_3|^{2n_3}\times
   \Tr\left[\frac{1}{i\cross{\partial}+\cross{\mu}_{j1}}e^{2\q{1}{a_1}\cdot(\rr-\vu_1(x))}
 \frac{1}{i\cross{\partial}-\cross{\mu}_{j2}}e^{-2\q{2}{a_1}\cdot(\rr-\vu_2(x))}
 \frac{1}{i\cross{\partial}+\cross{\mu}_{j3}}e^{2\q{1}{a_2}\cdot(\rr-\vu_1(x))}...\right]\;.
\end{equation}
We are interested in evaluating only the contribution from such
a diagram in which no derivatives act on any $\vu_I$ fields.  This
contribution is given by
\begin{equation}
\begin{split}
{\cal I}\propto
|\Delta_1|^{2n_1}|\Delta_2|^{2n_2}|\Delta_3|^{2n_3}\times\Tr\Biggl[&
\exp\Bigl\{i\sum_I 2\vu_I(x)\cdot(\sum_{a=a_1}^{a_{n_I}} \q{I}{a} -
\sum_{b=b_1}^{b_{n_I}}
\q{I}{b})\Bigr\}\\
&\times \frac{1}{i\cross{\partial}+\cross{\mu}_{j1}}e^{2\q{1}{a_1}\cdot\rr}
 \frac{1}{i\cross{\partial}-\cross{\mu}_{j2}}e^{-2\q{2}{a_1}\cdot\rr}
 \frac{1}{i\cross{\partial}+\cross{\mu}_{j3}}e^{2\q{1}{a_2}\cdot\rr}...\Biggr]
\label{separate u}\;.
\end{split}
\end{equation}
\end{widetext}
Momentum conservation implies that the net momentum added to the
loop is zero, i.e.
\begin{equation}
\sum_I \left(\sum_{a=a_1}^{a_{n_I}} \q{I}{a} -
\sum_{b=b_1}^{b_{n_I}}\q{I}{b}\right)=0\;.
\label{momentum cons}
\end{equation}
(As we argued after Eq.~(\ref{momentum conservation}), the momenta
contributed by $\phi$ is much smaller than $|\q{I}{}|$.)

We now recall from Section II that
the magnitude of the $\q{I}{}$ are different for different $I$.
($|\q{2}{}|=\eta\dm{2}$ is close in value to
$|\q{3}{}|=\eta\dm{3}$, but they
are not exactly equal because $\dm{2}$ and $\dm{3}$ differ
by terms of order $M_s^4/\mu^3$.)
This means that the momentum conservation condition (\ref{momentum cons})
can only be satisfied if
\begin{equation}
\sum_{a=a_1}^{a_{n_I}} \q{I}{a} -
\sum_{b=b_1}^{b_{n_I}}\q{I}{b}=0\;,
\end{equation}
separately for {\it{each}} $I$. This implies that in Eq.~(\ref{separate u}),
the coefficients of each of the $\vu_I$ cancel, and thus implies that
Eq.~(\ref{separate u}), the contribution of a generic one-loop diagram
in which no derivatives act on any $\vu_I$'s, is independent of $\vu_I$,
making it a trivial constant in the phonon effective action.  The phonons
are therefore massless to all orders in $\Delta$ and $\vu_I$.

As a special case, we can now demonstrate Eq.~(\ref{S00}) explicitly.
That is, we can show explicitly that
%\begin{equation}
$-\frac{1}{4G}\intspace{x} {\rm
tr}_{CF}\bigl((\Delta^u_{CF})^\dagger\Delta^u_{CF}\bigr)
$
%\end{equation}
is independent of $\vu_I$ as must be the case since it includes no
derivatives acting on $\vu_I$, and so would constitute a mass
term for the $\vu_I$ if it were to depend on the $\vu_I$. Indeed,
\begin{widetext}
\begin{equation}
\begin{split}
-\frac{1}{4G}\intspace{x} {\rm tr}_{CF}\bigl((\Delta^u_{CF})^\dagger\Delta^u_{CF}\bigr)
&=
-\frac{1}{G}\intspace{x}
\sum_I\bigl(\Delta_I\Delta^*_I\bigr)\sum_{\q{I}{a}}\sum_{\q{I}{b}}
e^{2\q{I}{a}\cdot(\rr-\vu_I(x))}
e^{-2\q{I}{b}\cdot(\rr-\vu_I(x))}\\
&=-(VT)\frac{1}{G}\sum_I(\Delta_I\Delta^*_I)P_I\; ,
\end{split}
\end{equation}
\end{widetext}
as given in (\ref{S00}).

\section{Single plane wave \label{single pw}}

In this Appendix, we investigate phonons in the presence of a condensate for
which only one of the three $\Delta_I$ is nonzero, meaning that
only quarks with two different colors and flavors form Cooper pairs.
We take $\Delta_3\neq 0$ and $\Delta_1=\Delta_2=0$. This implies
that there is pairing only between $ur$ and $dg$ quarks, and between
$ug$ and $dr$ quarks, where $r,g$ and $b$ refer to the colors of the
quarks. Furthermore, we assume that the set $\setq{3}{}$ contains
only one vector, $\bf q$, so that $\Delta_3(x)$ varies in space as a
single plane wave, $\Delta_3(\rr)=\Delta_3\exp(2i\q{}{}\cdot\rr)$.
In this simple case it is possible to derive the phonon effective
action and the shear moduli without employing the Ginzburg-Landau
expansion, working to order $\phi^2$.

The oscillation of the condensate is described by a single  phonon
field, $\vu_3(x)$. In analogy with Eq.~(\ref{cond phon1}), we define
%\begin{widetext}
\begin{eqnarray}
\Delta_3^u(x)&=&\Delta_3e^{2i\q{}{}\cdot(\rr-\vu_3(x))}\nonumber\\
&\approx&
\Delta_3e^{2i\q{}{}\cdot\rr}\left(1-i\phi(x) - \ha \phi(x)^2\right)
\end{eqnarray}
%\end{widetext}
where
$\phi(x)=2\q{}{}\cdot\vu_3(x)$.

This condensate breaks translational symmetry in the
$\unitq{}{}$ direction, but leaves an $O(2)$
symmetry corresponding to rotations about the $\unitq{}{}$-axis
unbroken. Taking $\unitq{}{}$ along the $z$ axis, the potential
 energy ${\cal{U}}$ of Eqs.~(\ref{general pe}) and (\ref{diagonal pe})
 must be symmetric under  rotations about  the $z$-axis, taking the form
\begin{equation}
\begin{split}
{\cal{U}}=& \frac{\lambda^{zzzz}}{2} (s^{zz})^2\\
& + \frac{\lambda^{xzxz}}{2} \left( (s^{xz})^2 + (s^{zx})^2+(s^{yz})^2+(s^{zy})^2
 \right) \label{U from symmetry}\,,
\end{split}
\end{equation}
where the strain tensors $s^{mu}$ are defined in (\ref{strain}).
$\lambda^{zzzz}$ and $\lambda^{xzxz}=\lambda^{yzyz}$ are the two independent
elastic moduli that we will now evaluate.
% from the microscopic
%theory.

In writing the Lagrangian in terms of the Nambu-Gorkov fields, we
ignore the quarks that do not participate in pairing. We also note
that the inverse propagator can be written as a block diagonal
matrix made up of four blocks that correspond to the $ur$ particles
and $dg$ holes, $dg$ particles and $ur$ holes, $ug$ particles and
$dr$ holes and $dr$ particles and $ug$ holes. Since only quarks
that belong to the same block interact, the inverse Nambu-Gorkov
propagator can  be written as
\begin{equation}
S^{-1} = S^{-1}_{ur-dg}\oplus S^{-1}_{dg-ur}\oplus S^{-1}_{ug-dr}\oplus S^{-1}_{dr-ug}\label{pw
Sinv}\;,
\end{equation}
with
\begin{widetext}
\begin{equation}
\begin{split}
&S^{-1}_{ur-dg}=\left( \begin{array}{cc}
i\cross{\partial}+\cross{\mu}_{u} & (C\gamma^5)\Delta_3^u(x)    \\
-(C\gamma^5)\Delta_3^{u*}(x) &
(i\cross{\partial}-\cross{\mu}_{d})^T
\end{array} \right)\;,\;
S^{-1}_{dg-ur}=\left( \begin{array}{cc}
i\cross{\partial}+\cross{\mu}_{d} & (C\gamma^5)\Delta_3^u(x)    \\
-(C\gamma^5)\Delta_3^{u*}(x) &
(i\cross{\partial}-\cross{\mu}_{u})^T
\end{array} \right)\;,\;\\
&S^{-1}_{ug-dr}=\left( \begin{array}{cc}
i\cross{\partial}+\cross{\mu}_{u} & -(C\gamma^5)\Delta_3^u(x)    \\
(C\gamma^5)\Delta_3^{u*}(x) &
(i\cross{\partial}-\cross{\mu}_{d})^T
\end{array} \right)\;,\;
S^{-1}_{dr-ug}=\left( \begin{array}{cc}
i\cross{\partial}+\cross{\mu}_{d} & -(C\gamma^5)\Delta_3^u(x)    \\
(C\gamma^5)\Delta_3^{u*}(x) &
(i\cross{\partial}-\cross{\mu}_{u})^T
\end{array} \right)\label{Sinv blocks}\;.
\end{split}
\end{equation}
\end{widetext}
The phonon effective action is obtained by integrating out the Nambu-Gorkov fields,
yielding the result
\begin{equation}
\begin{split}
i{\cal{S}}[\vu]=\log(Z[\vu])=&-i\frac{1}{G}(VT)|\Delta_3|^2 \\
 &+ 2
 {\rm Tr}_{{\rm ng}}\log\left(S^{-1}_{ur-dg}\right)\label{Z for pw}\;,
\end{split}
\end{equation}
where we have used the property that the contributions from the four
blocks, $S^{-1}_{ur-dg}$, $S^{-1}_{ug-dr}$, $S^{-1}_{dr-ug}$ and
$S^{-1}_{dg-ur}$, are equal. Since the trace of an operator does not
change upon making a unitary transformation, corresponding to a
change of basis, we can simplify $S^{-1}$ by choosing a basis
which gets rid of the
$\exp(2i\q{}{}\cdot\rr)$ and the $C\gamma^5$ appearing with
$\Delta_3$. In the new basis,
\begin{widetext}
\begin{equation}
\begin{split}
S^{-1}_{ur-dg} = &  \left( \begin{array}{cc}
e^{-i\q{}{}\cdot\rr} & 0   \\
0 &  (C\gamma^5)e^{i\q{}{}\cdot\rr} \end{array} \right)
\left( \begin{array}{cc}
i\cross{\partial}+\cross{\mu}_{u} & (C\gamma^5)\Delta_3^u(x)    \\
-(C\gamma^5)\Delta_3^{u*}(x) &
(i\cross{\partial}-\cross{\mu}_{d})^T
\end{array} \right)
\left( \begin{array}{cc}
e^{i\q{}{}\cdot\rr} & 0   \\
0 &  (-C\gamma^5)e^{-i\q{}{}\cdot\rr} \end{array} \right)\\
&=\left( \begin{array}{cc}
i\cross{\partial}+\cross{q}+\cross{\mu}_{u} &
     \Delta_3(1-i\phi-\ha\phi^2)    \\
\Delta_3^*(1+i\phi-\ha\phi^2) &
i\cross{\partial}-\cross{q}-\cross{\mu}_{d}
\end{array} \right)\ .
\end{split}
\end{equation}
\end{widetext}

In this simple case, the inverse propagator in the absence of the
phonons can be inverted. (This is why we do not need to resort to
the  Ginzburg Landau approximation.) Hence we separate
$S^{-1}_{ur-dg}$ into $S^{-1}_{ur-dg}=S^{-1}_{\Delta}+\Sigma$ with
\begin{equation}
S^{-1}_{\Delta}=\left( \begin{array}{cc}
i\cross{\partial}+\cross{q}+\cross{\mu}_{u} & \Delta_3  \\
\Delta_3^* &   (i\cross{\partial}-\cross{q}-\cross{\mu}_{d})
\end{array} \right)\;
\end{equation}
and
\begin{equation}
\Sigma=\left( \begin{array}{cc}
0 & \Delta_3(-i\phi-\ha\phi^2)    \\
\Delta_3^*(+i\phi-\ha\phi^2) &  0
\end{array} \right)\,.
\end{equation}
Since we have rotated out the phase $\exp(2i\q{}{}\cdot\rr)$ from
$\Delta_3$, the off-diagonal components of $S^{-1}_{\Delta}$ do not
depend on position anymore.  Upon inverting  $S^{-1}_{\Delta}$, one gets
the full propagator for the fermions in the absence of phonons, to
all orders in $\Delta$.  This propagator is diagonal in momentum space and
therefore can be written in a simple way employing the weak-coupling
approximation and the HDET
formalism (implemented as in Eqs. (\ref{prop1}) and (\ref{prop2}) or more formally as
in Ref.~\cite{Nardulli:2002review}), yielding
\begin{widetext}
\begin{equation}
S_{\Delta}(p)= \frac{1}{D_{\Delta}(p)}\left( \begin{array}{cc}
(\tilv\cdot p-\vv\cdot\q{}{}-\delta\mu_3) & -\Delta_3  \\
-\Delta_3^* &  (V\cdot p-\vv\cdot\q{}{}-\delta\mu_3)
\end{array} \right)\;,
\label{Sdelta}\end{equation}
\end{widetext}
where \bea D_{\Delta}(p) &=& (\tilv\cdot
p-\vv\cdot\q{}{}-\delta\mu_3)(V\cdot p-\vv\cdot\q{}{}-\delta\mu_3)\nonumber \\
& &-|\Delta_3|^2 \,. \label{Ddelta}\eea

\begin{center}
\begin{figure*}[t]
\includegraphics[width=5.in,angle=0]{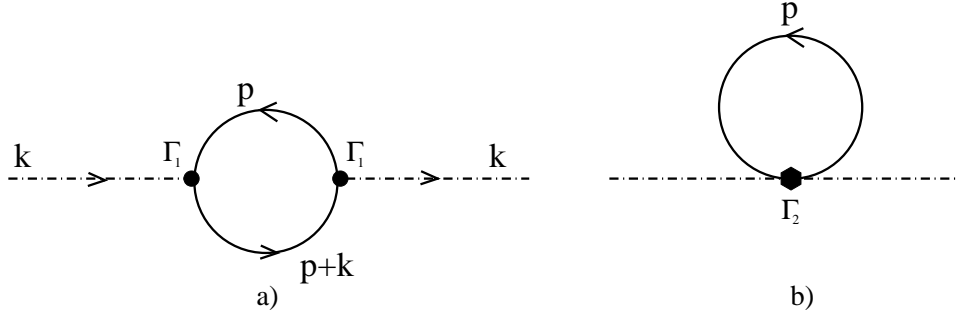}
\caption{ Diagrams contributing to the phonon self-energy. The
dot-dashed lines correspond to the propagator of the phonon fields.
Full lines correspond to the full propagator of quark
quasi-particles. The phonon-quark-quark vertices $\Gamma_1$ and the
phonon-phonon-quark-quark vertex $\Gamma_2$ are also shown. Diagram
b$)$ contributes only to the zero momentum polarization tensor.
Diagram a$)$ is nonvanishing for any value of the external momentum
$k$ of the phonon.} \label{phononself}
\end{figure*}
\end{center}

We now consider the interaction with the phonon field. To second
order in $\phi$, there are two interaction vertices,
$i\phi(x)\Gamma_1$ and $-\ha\phi(x)^2\Gamma_2$, where
\begin{equation}
\Gamma_1=\left( \begin{array}{cc}
0 & -\Delta_3    \\
\Delta_3^* & 0
\end{array} \right) {\mbox{ and }}
\Gamma_2=\left( \begin{array}{cc}
0 & \Delta_3    \\
\Delta_3^* & 0
\end{array} \right) \,.
\end{equation}
The phonon effective action to order $\phi^2$ is given by
the diagrams in Fig.~\ref{phononself}, namely
\be
{\cal S}^{\phi^2}  = {\rm Tr}(S_{\Delta}\,
\phi\, \Gamma_1  S_{\Delta}\, \phi\, \Gamma_1)  - {\rm Tr}(S_{\Delta}\,
\phi^2\, \Gamma_2)\, ,  \label{actionfonon}
\ee
where the trace is over
Nambu-Gorkov indices and space-time.
Evaluating the action along the same lines
as in Section \ref{simplifying piak}, we find
\begin{widetext}
 \be {\cal
S}^{\phi^2}  =  \frac{\mu^2}{ \pi^2} \Delta_3^2 \int \frac{d^4
k}{(2\pi)^4} \phi(k)\phi(-k) \intpos\intv
 \frac{(V\cdot k)( \tilde V\cdot
k)}{D_{\Delta}(p+k)D_{\Delta}(p)} \label{actionfonon2}\,, \ee
%\end{widetext}
where the $k$-independent
second diagram in Fig.~\ref{phononself}
has cancelled the $k$-independent
contribution from the first diagram.
We then integrate over $p_0$ and ${\hat {\bf v}}$ analytically
and do the $s$-integral numerically.  Upon returning
to position space, the action takes the form (\ref{full action}) which, in
this simple case, reduces to
%\begin{widetext}
\begin{equation}
{\cal{S}}^{\phi^2}=\ha\intspace{x}\left[\frac{\rho}{4q^2} (\partial_0 \phi(x))^2
- \frac{\kappa_{\bot}}{4q^2} (\partial_{\bot} \phi(x))^2  - \frac{\kappa_{||}}{4q^2}
(\partial_{||} \phi(x))^2 \right]  \label{Seff pw}\;,
\end{equation}
\end{widetext}
where
%\begin{equation}
$\partial_{||} \equiv\unitq{}{}(\unitq{}{}\cdot\vec{\partial})$ and
%{\mbox{  and }}
$\partial_{\bot} \equiv \vec{\partial}-\partial_{||}$.
%\end{equation}
We have written the  factors of $4q^2=4\eta^2\delta\mu^2$
in the denominators in (\ref{Seff pw}) in order to facilitate
comparison to (\ref{U from symmetry}).
The potential energy ${\cal{U}}$ in (\ref{Seff pw})
can easily be written in the form
given in Eq.~(\ref{U from symmetry}) by substituting $\phi=2{\bf q}\cdot {\bf u}_3$
and keeping only the term symmetric in
$\partial_i \vu_j$. Taking $\unitq{}{}$ in the $z$ direction, one
can easily see that $\lambda^{zzzz}=\lambda_{||}$ and
$\lambda^{xzxz}=\lambda^{yzyz}=\lambda_{\bot}$.

The coefficients $\rho$, $\lambda_{\perp}$ and
$\lambda_{\parallel}$ are each proportional to $\mu^2$ and
have nontrivial $\Delta_3$- and $\delta\mu_3$-dependence.
We have obtained them numerically for arbitrary values of $\Delta_3$,
and by plotting them versus $\Delta_3^2$ at small $\Delta_3$ we have
checked that
$\rho$ and $\lambda_{||}$ are proportional to $\Delta_3^2$ and $\lambda_{\bot}$
is consistent with being proportional to $\Delta_3^4$.  Within the accuracy of our
numerical analysis, we find $\rho=\lambda_{||}=\kappa$ to order $\Delta_3^2$,
where $\kappa$ is given by (\ref{kappavalue}).
This is in agreement with what we obtain from the phonon effective action
(\ref{Seff4}) evaluated to order $\Delta_3^2$ in the Ginzburg-Landau approximation,
upon specializing (\ref{Seff4}) to the single plane wave case. It therefore
provides a useful check on our calculations.
Whether via the Ginzburg-Landau analysis of the main Sections of this
paper or via the numerical all-order-in-$\Delta_3$ evaluation of this Appendix,
we learn that  $\lambda^{zzzz} =\kappa$ and
$\lambda^{yzyz}=0$ to order $\Delta_3^2$, from which we conclude that the
shear modulus of this single-plane-wave ``crystal'' is zero, to this order, and
is presumably proportional to $\mu^2\Delta_3^4/\delta\mu^2$.

\end{document}